\documentclass[sigconf]{acmart}

\AtBeginDocument{%
  \providecommand\BibTeX{{%
    \normalfont B\kern-0.5em{\scshape i\kern-0.25em b}\kern-0.8em\TeX}}}

\copyrightyear{2021}
\acmYear{2021}
\acmConference[Conference '21]{Conference '21}{}{}
\acmBooktitle{Conference '21}
\usepackage{microtype}
\usepackage{float}
\usepackage{nicefrac}       % compact symbols for 1/2, etc.
\usepackage{microtype}      % microtypography
\usepackage{graphicx}
\usepackage{amsfonts}       % blackboard math symbols
\usepackage{amsmath}
\usepackage{amsthm}

\usepackage{bm}
\usepackage{bbm}
\usepackage{booktabs}
\usepackage{caption}
\usepackage{colortbl}
\usepackage{color}
\usepackage{hyperref}

\usepackage{natbib}
\usepackage{enumitem}
\usepackage[most]{tcolorbox}
\usepackage{thmtools,thm-restate}

\usepackage{todonotes}
\usepackage[algoruled,boxed,lined]{algorithm2e}
\newtheorem{prop}{Proposition}

\newtheorem*{prop*}{Proposition}
\newtheorem*{claim*}{Claim}
\newtheorem*{cor*}{Corollary}

 \definecolor{Gray}{gray}{0.9}

\begin{document}

\title{\textit{Bayesian Consensus}: Consensus Estimates from Miscalibrated Instruments under Heteroscedastic Noise}


\author{Chirag Nagpal$^{1,2}$, Robert E. Tillman$^{1}$, Prashant P. Reddy$^{1}$, Manuela M. Veloso$^{1}$\\
{\large  $^{1}$J.P. Morgan AI Research\\ $^{2}$Carnegie Mellon University}}

% \author{Chirag Nagpal$^{1,2}$, Robert E. Tillman$^{1}$, Prashant P. Reddy$^{1}$, Manuela M. Veloso$^{1}$}
% \affiliation{%
%  \institution{$^{1}$J.P. Morgan AI Research\\ $^{2}$Carnegie Mellon University}
%  \email{chiragn@cs.cmu.edu, robert.tillman}
%  \country{ }
% }

\renewcommand{\shortauthors}{JPMorgan AI Research}

\begin{abstract}
We consider the problem of aggregating predictions or measurements from a set of human forecasters, models, sensors or other instruments which may be subject to bias or miscalibration and random heteroscedastic noise. We propose a Bayesian consensus estimator that adjusts for miscalibration and noise and show that this estimator is unbiased and asymptotically more efficient than naive alternatives. We further propose a Hierarchical Bayesian Model that leverages our proposed estimator and apply it to two real world forecasting challenges that require consensus estimates from error prone individual estimates: forecasting influenza like illness (ILI) weekly percentages and forecasting annual earnings of public companies. We demonstrate that our approach is effective at mitigating bias and error and results in more accurate forecasts than existing consensus models.
\end{abstract}






\maketitle

\section{Introduction}

\emph{Forecasting} refers to the science of predicting future events and quantities. Forecasts are often relied on to inform important decisions in public policy, healthcare, economics, finance and other fields \citep{ernst:16}. The value of accurate predictions and individuals, models and other instruments that can accurately predict future events and quantities has led governments and private companies to invest in identifying individuals who are uniquely skilled in making forecasts \citep{ungar:12}.

Predictions from multiple individuals and models are often aggregated to produce more accurate \emph{consensus} forecasts. The literature is nearly unanimous in finding that consensus predictions are more accurate than individual predictions \citep{vandijk:19}. How to best aggregate predictions, however, is largely an open problem: some empirical studies argue for giving increased weight to individuals or instruments that have historically been more accurate \citep{ungar:12}, while other studies conclude it is difficult to outperform a simple uniform average \citep{graefe:11}.

While much of the forecasting literature focuses on scenarios where forecasts are generated by humans, the problem of aggregating predictions remains relevant when predictions come from other instruments such as models from statistics, machine learning, epidemiology and econometrics, prediction markets, exchanges and other non-human sources.

A related problem, \emph{Blind Calibration} \citep{balzano:07, gribonval2012blind}, refers to aggregating measurements, as opposed to forecasts, from noisy sensors, which are potentially \emph{miscalibrated} and demonstrate systematic bias. Our problem differs from this setting in that since we work with forecasts, we have access to historical errors to adjust our estimates.

In the sections that follow, we make following contributions:

\begin{itemize}[leftmargin=1em]
    \item We consider the general problem of aggregating sets of predictions of $k$ independent numerical quantities from $n$ instruments and propose a Bayesian estimator to recover the underlying quantities.
    \item We analyze the theoretical properties of the proposed estimator in comparison to other alternatives. We show that it is unbiased and provide conditions under which it is efficient.
    \item We propose a Latent Variable model motivated by our estimator and apply it to real world forecasting challenges for percentages of influenza like illnesses and future earnings of public companies.  We demonstrate that our approach mitigates forecast bias and error and results in more accurate consensus forecasts than the uniform average and other consensus estimators based on regression.
\end{itemize}

\begin{figure*}[!htbp]
    \begin{minipage}{0.475\textwidth}
    \centering
    \includegraphics[width=0.75\linewidth]{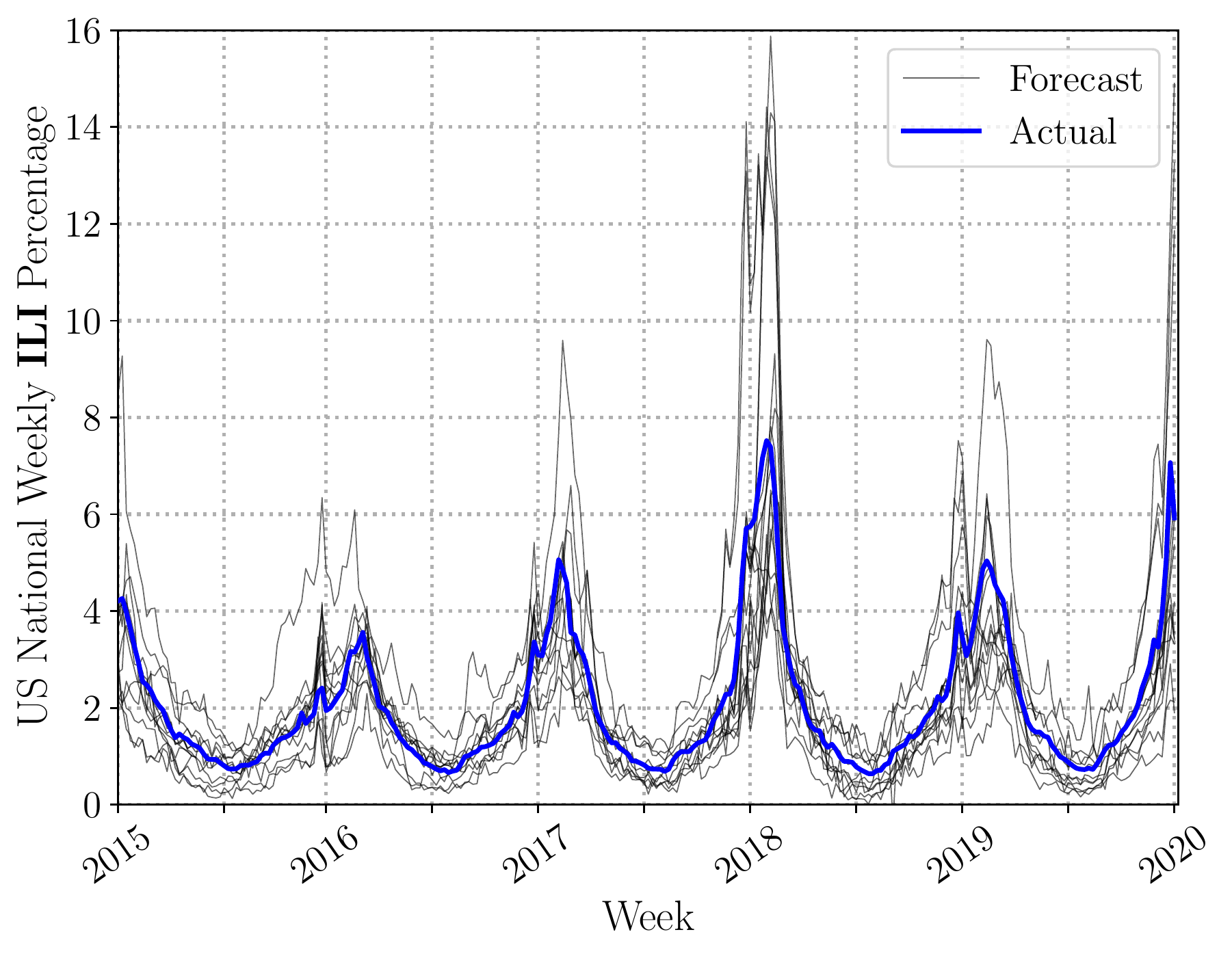}
    \captionof{figure}{ILI Forecasts of individual sensors along with actual percentages over a 5 year period. Notice that some sensors consistently over/underestimate the ILI percentages. }
    \end{minipage}\hspace{1em}
    \begin{minipage}{0.475\textwidth}
    \centering
    \includegraphics[width=0.75\linewidth]{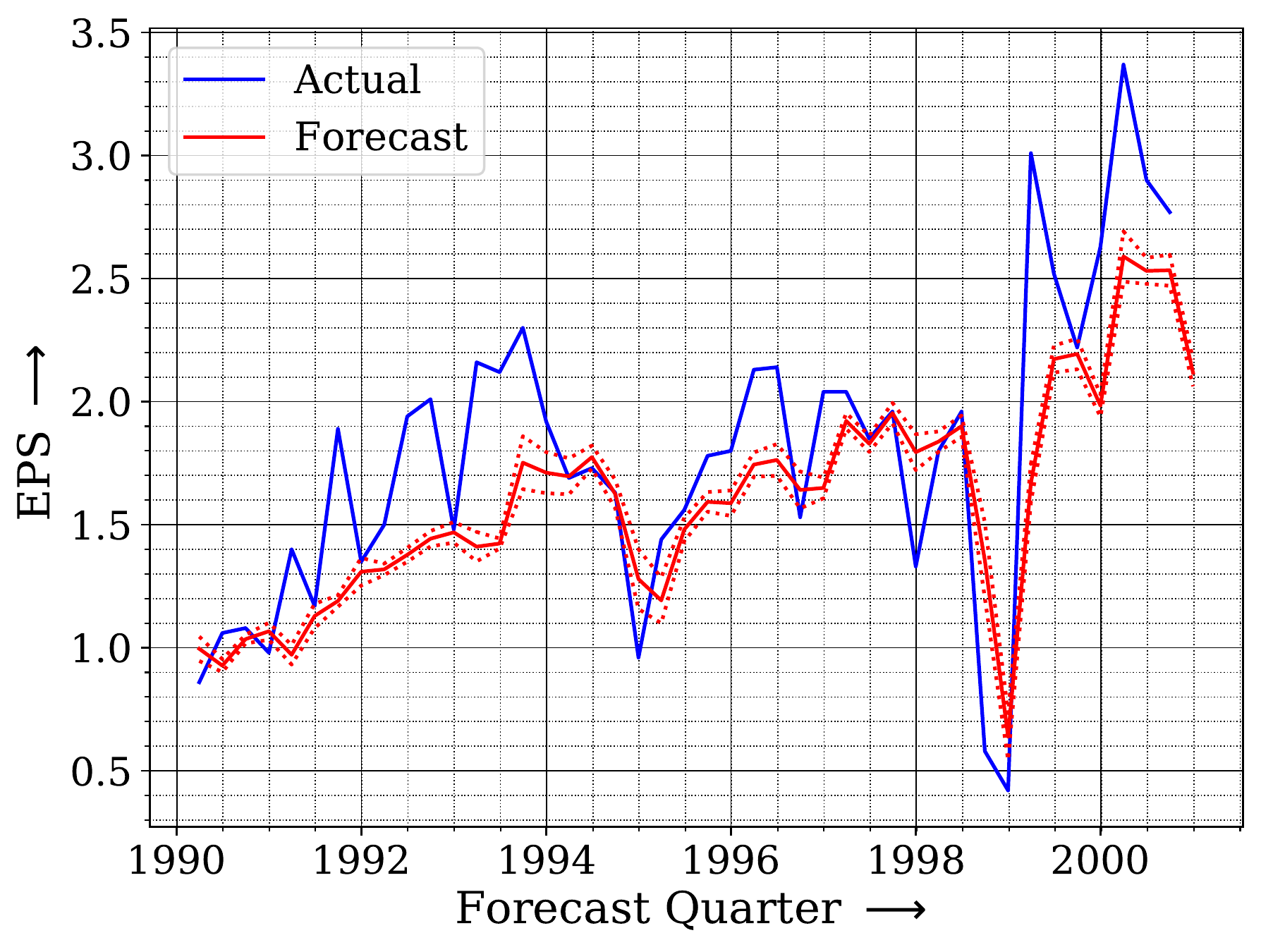}
    \captionof{figure}{Consensus earnings estimate versus actual earnings for a large US financial firm over a 10 year period.  }
    \end{minipage}
    \label{fig:example}
\end{figure*}

\section{Motivations}

\textbf{Influenza Like Illness (ILI)  Forecasting:} The term \emph{ILI} refers collectively to influenza, the common cold and other flu-like respiratory illnesses with symptoms such as fever, dry cough, nausea and body aches. In the light of the 2019 Novel Coronavirus (COVID-19) pandemic, the US Centre for Disease Control and  Prevention (CDC) has expanded the use of ILI to include COVID-19. The CDC carries out monitoring of the spread of ILI at state and national levels and releases patient visits data every week. The ability to monitor and forecast the growth of ILI is hence of vital importance to public officials and healthcare professionals as it provides actionable information to make decisions about public policy interventions like social distancing and to better triage healthcare resources like personal protective equipment and critical care equipment for life support. 

ILI forecasting has thus received increased attention recently. Most forecasting methods are based on autoregressive models trained using multiple data sources including individual patient visits, state level visits, drug sales and, more recently, social media activity. In the existing literature, these are referred to as \emph{sensors} and averaged to produce a consensus forecast of ILI percentages \citep{brooks2015flexible, farrow2017human,jahja:2019,farrow:2016}. The CDC uses a consensus estimate as its primary forecast \citep{reich:19}.

 \citep{maves:20}. Accurate forecasts of the severity and timing of outbreaks in different regions allow for a coordinated response where resources are moved between high impact region. Such a response, however, requires buy-in from multiple affected parties which is unlikely unless forecasts are reliable and trusted. The CDC is currently tracking COVID-19 forecasts from 14 different research groups, some of which vary significantly. While there is value in having multiple types of forecasts, delays and uncertainty also result as officials weigh the pros and cons of individual models forecasting different outcomes. 

Similarly, as governments consider implementing policies like social distancing and quarantines to reduce the severity of outbreaks and prevent medicals systems from exceeding capacity versus the broader economic, psychological and other societal impacts of those policies, many officials have delayed implementing such policies while waiting on additional data to confirm the accuracy of forecasts. Even small delays, however, can significantly affect the scale of an outbreak and the effectiveness of policy interventions \citep{matrajt:20}. A robust, trusted methodology for forecast consensus can help reduce the uncertainty public officials face when making these difficult decisions allowing for more timely interventions and improved health outcomes.

Finally, the COVID-19 crisis has given rise to many amateur forecasters publishing their own forecasts through blogs and social media. Many of these forecasters do not any background in public health or epidemiology; some of their forecasts have severe methodological flaws. The overwhelming number of available forecasts, professional and amateur, has led to considerable confusion among the public regarding how their  community may be affected and how much precaution they should take. This confusion has also created opportunities for public officials and others to rely on and publicize models which forecast a desired outcome rather than choosing models based on accuracy or methodological soundness. Our consensus model provides a solution to such uncertainty by aggregating over any number of individual forecasts while (unlike some other consensus models) being robust to noisy and potentially unsound forecasts through our latent group assignment. 

These impacts apply not just to COVID-19, but other future seasonal and pandemic flus. The CDC recently moved from using a simple uniform consensus model for seasonal flu forecasting to an weighted model for the 2018-2019 flu season \citep{reich:19}. Our experiments have shown that our LVBC consensus model is more accurate than simple  weighted consensus models; we thus believe it will lead to improved forecasting of future ILI with the above impacts to public health.

\noindent \textbf{Earnings Forecasting for Public Companies:} The \emph{earnings forecasting} problem refers to predicting future earnings of public companies at various horizons, e.g. revenue or net profit over the next quarter or year. Analysts at financial institutions regularly issue such forecasts for companies in their coverage universe, typically referred to as \emph{earnings estimates}. Investors and other market participants often rely on the \emph{consensus earnings estimate}, or the simple uniform average of the individual estimates produced by analysts, to make investment decisions and value companies. While the literature has consistently found analysts' earnings estimates to be more accurate, at the aggregate level as compared to autoregressive models trained on historical data \citep{brown78,fried:82} and found a strong relationship between analysts’ forecast revisions and changes in companies' stock prices \citep{michaely05}. However, there is also significant empirical evidence of systematic bias and error in individual analyst estimates in finance and behavioral economics communities.

Much of this literature focuses on over-optimism in analysts’ forecasts \citep{debondt90, hong03, elliot10} which is often explained as resulting from incentives for analysts to maintain good relations with companies’ management \citep{francis93, richardson99}. Other literature finds biases due to possible conflicts of interest with companies who are also investment banking clients \citep{michaely99, lin05}. Finally, recent research has found evidence of systematic racial, gender and political biases in analysts' earnings forecasts, i.e. analysts' forecasts are more optimistic for companies with "in-group" CEOs \citep{jannati19}.

Research also shows that biases are stronger in some analysts than others \citep{michaely99, hong00}, which supports the use of a consensus model like LVBC since it can assign more accurate and less biased analysts to different latent groups. Many large investors rely on the simple consensus earnings forecast to guide their investment decisions. There is thus value is the adoption of a consensus model like LVBC which can mitigate individual analysts' biases through latent group assignment and adjustments so that analyst biases due not affect which companies receive investment capital. Mitigating such biases impacts both the efficient allocation of capital to successful companies, promoting orderly and efficient markets, and ensuring that companies founded by underrepresented groups are not systematically less likely to receive funding due to racial, gender or other biases reflected in the simple consensus average of analyst estimates.

Such bias, unaccounted for, can result in over(under)-inflation of company valuations and the investments they receive as a result of conflicts of interest \citep{michaely99, lin05} or for discriminatory reasons \citep{jannati19}. There is thus value in a consensus estimator that is robust to systematic bias and error in analyst earnings estimates.

\section{Analysis}
\label{sec:analysis}
We now state formally the consensus estimation problem along with the assumptions we make about the data generating process and notation. We introduce estimators for the predicted or measured quantity of interest (QoI) and demonstrate the conditions under which our proposed \textit{Bayesian Consensus} estimator outperforms other consensus estimators.

\subsection{The Formal Setting}

In our problem setting, we would like to recover a scalar quantity $X \in \mathbb{R}$, but only have access to noisy measurements of this quantity. For simplicity, we assume there are two classes of measurements, both subject to white noise of different intensities. Out of these two classes, one set of measurements are \textit{good}, in the sense that they are unbiased with mean $\mu$ and variance $\sigma^{2}$, while the other  are \textit{bad} in the sense that they are biased with mean ($\alpha\cdot\mu + \beta$), where $\alpha, \beta \in \mathbb{R}$. The \textit{bad} instruments are subject to a different random error $\sigma_*^{2}$ as compared to the \textit{good} measurements. The class membership of each measurement is indicated by a binary vector $\mathbf{Z} \in \{0,1 \}^{m+n}$. We denote the \textit{good} measurements as $\hat{X}_i$ and the \textit{bad} measurements as ${X}^*_j$. 
% \begin{align*}
% \text{Thus,} \quad \mathbb{E}[\hat{X}_i] &= \mu,  &&\text{Var}[\hat{X}_i] = \sigma^{2} \\
% \quad \mathbb{E}[X^{*}_j] &= \alpha\cdot\mu +\beta, &&\text{Var}[\hat{X}_j] = \sigma_*^{2} 
% \end{align*}
Thus, 
\begin{align*}
\mathbb{E}[\hat{X}_i] = \mu, \; \; \text{Var}[\hat{X}_i] = \sigma^{2} \quad \text{and,} \; \;  \mathbb{E}[X^{*}_j] = \alpha\cdot\mu +\beta, \; \; \text{Var}[\hat{X}_j] = \sigma_*^{2}
\end{align*}

The total number of \textit{good} measurements is $m$ while the total number of \textit{bad} measurements is $n$. The set of all measurements is notated jointly as $\mathcal{D} =  \{ \hat{X}_i \}_{i=1}^{m} \cup \{ {X}^{*}_j \}_{j=1}^{n}$

\subsection{Estimators for $X$}
\label{sec:estimators}
We now introduce some estimators for $X$ and study their properties in terms of bias and variance. We then propose a Bayesian approach to recover $X$ and demonstrate its superiority over other estimators. Note that the estimators introduced are functions of the data, $\mathcal{D}$.

\noindent  \textbf{Naive Estimator ($f_\text{NE}$)}: This is the simple uniform average over  all the measurements. $$ f_{\text{NE}}(\mathcal{D}) = \frac{\hat{X}_1 + \hat{X}_2 + ... + \hat{X}_m +  {X}^*_1 + {X}^*_2 + ... + {X}^*_n}{m+n} $$ 
\vspace{-1em}    
\noindent \begin{restatable}{prop}{one}{The \textbf{Naive Estimator} {\normalfont($f_{\text{NE}}$)} is asymptotically \textbf{biased} in $n$.}
    \label{prop:one}
    \end{restatable}
    
   \noindent Proof Sketch. Immediately from $\lim\limits_{n\to\infty} \text{Bias}(f_{\text{NE}}) \neq 0$. \hfill \hfill \hfill\hfill\hfill\hfill\hfill\hfill\hfill\hfill\hfill\hfill  \hfill \hfill \hfill\hfill\hfill\hfill\hfill\hfill\hfill\hfill\hfill\hfill (Appx. \ref{apx:one}) $\hfill \blacksquare$
    
    \noindent \textbf{Conservative Estimator} ($f_\text{CE}$): Assuming we have knowledge of $Z$, i.e. we know which instruments are \textit{good} or \textit{bad}, we can improve on the Naive Estimator by excluding the \textit{bad} instruments. 
 $$f_{\text{CE}} = \frac{\hat{X}_1 + \hat{X}_2 + ... + \hat{X}_m}{m}$$
    % $$\text{Bias}(f_{\text{CE}} ) =  \mathbb{E}[f_{\text{CE} }] - \mu = 0, \quad \text{Var}[f_{\text{CE}}]  =\frac{\sigma^2}{m} $$ 
    
 \begin{restatable}{prop}{two}{The \textbf{Conservative Estimator} {\normalfont($f_{\text{CE}}$)} is \textbf{unbiased} in $m$.} 
    \label{prop:two}
    \end{restatable}
    
    \noindent Proof Sketch. Immediately from $  \lim\limits_{m\to\infty} \text{Bias}(f_{\text{CE}}) = 0$. \hfill\hfill\hfill\hfill\hfill\hfill\hfill\hfill\hfill\hfill\hfill \hfill \hfill\hfill\hfill\hfill\hfill\hfill\hfill\hfill\hfill\hfill(Appx. \ref{apx:two}) $\hfill \blacksquare$
  
    \noindent \textbf{Greedy Estimator} ($f_\text{GE}$):   Now, let us assume we have access to $\mathbf{Z}$, as well as the true coefficients of the linear miscalibration, $\alpha$ and $\beta$. In this case we can recover an unbiased estimate of $X$ as follows.
     $$ f_{\text{GE}}(\mathcal{D}) = \frac{  \sum\limits_{i=1}^{m} \hat{X}_i  + \frac{1}{\alpha}\left( \sum\limits_{j=1}^{n} {X}^*_i-n\beta\right)}{m+n} $$ 

\begin{restatable}{prop}{three}
The \textbf{Greedy Estimator}  {\normalfont($f_{\text{GE}}$)} is \textbf{unbiased} in $n$ and $m$.
\label{prop:three}
\end{restatable}
\noindent Proof Sketch. Immediately from
$$\hspace{3em} \lim\limits_{m\to\infty}  \text{Bias}(f_\text{GE}) =  \lim\limits_{n\to\infty}  \text{Bias}(f_\text{GE}) = 0. \hspace{3em} \text{(Appx.  \ref{apx:three})} \blacksquare  $$ 
    % $\text{Now, }\quad\quad \quad \text{Var}(f_{\text{GE}}) &= \Big( \frac{1}{m+n} \Big)^2 \big[ m\sigma^2 + \frac{n \sigma_*^2}{\alpha^2}  \big] $

The natural question that arises is under what conditions is the greedy approach superior to the conservative estimator. We thus proceed to compare their efficiencies. Since both are unbiased estimators, comparing the MSE reduces to comparing their variances.

\begin{restatable}{prop}{gevsce} {The {\normalfont$ \text{MSE}(f_{\text{GE}}) \leq \text{MSE}(f_{\text{CE}})$} iff. $$\nicefrac{\sigma_*^2}{\sigma^2} \leq \big(\nicefrac{n}{m} + 2 \big) \alpha^2 .$$ \label{prop:gevsce}}
\end{restatable}
\vspace{-1em}
\noindent Proof Sketch. Set $\text{Var}(f_{\text{GE}})\leq \text{Var}(f_{\text{CE}})$, and the positivity of $\sigma^2$, $\sigma_*^2$, $n$ and  $m$. \hfill \hfill \hfill\hfill\hfill\hfill (Appx. \ref{apx:gevsce}) $\blacksquare$

% {\color{red}
% \begin{prop}For $|\alpha| > 1$, the \textbf{Greedy Estimator} has lower MSE than \textbf{Naive Estimator}. 
% \end{prop}

% Proof.   For $|\alpha| > 1$, $\text{Var}(f_\text{GA}) < \text{Var}(f_\text{NE})$,  and,
% $\text{MSE}(f) = \text{Bias}^2(f)+ \text{Var}(f)$
% $$\therefore  \qquad \text{MSE}(f_\text{GA}) \leq \text{MSE}(f_\text{NE}) \qquad \qquad \qquad  \qquad  \blacksquare$$

% \begin{prop}
% For $|\alpha| < 1$, the \textbf{Greedy Estimator} has lower MSE than \textbf{Naive Estimator} if and only if 
% $$\sigma_*^{2} \leq \frac{n \alpha^2 \big(\alpha-1)\mu + \beta\big)^2}{1-\alpha^2}$$ \label{prop:savsna}
% \end{prop}
% \vspace{-1em}
% Proof Sketch.  The above is a direct implication of $$\text{Var}(f_{\text{GE}})  \leq  \text{Var}(f_{\text{NE}} ) + \text{Bias}^2(f_\text{NE}).$$ Full proof is in Appendix \ref{appx:} $\qquad \qquad\qquad \qquad\qquad \blacksquare$ 
% }
% Proposition \ref{prop:savsna} is powerful in the sense that it states that the the relative efficiency of the \textbf{Greedy Estimator} estimator is characterized by only the \textit{bad} instruments. At the same time, Proposition \ref{prop:savsna} also demonstrates that the \textbf{Greedy Estimator} estimator is \textbf{NOT} arbitrarily better than \textbf{Naive Estimator}. In situations where instruments are likely to underestimate $X$, typically one would expect $|\alpha|<1$. Thus, in such scenarios, unless we have a strong reason to believe that $\sigma_*^2$ is small, the \textbf{Naive Estimator} may result in a more efficient estimator. 

\textbf{Bayesian Estimator ($f_\text{BE}$): } Let us assume that the quantity to be recovered, $X$, is random and place a weak normal prior on $X$. Since we do not know the individual variances of the instruments and hence for estimation we assume homoscedasticity with unit variance, we get the following estimator.
\begin{align*}
f_{\text{BA}}(\mathcal{D}) &= \frac{1}{(m+n\alpha^2 + \lambda_0) }\bigg(\sum_{i=1 }^{m} \hat{X}_i
 + \alpha \sum_{j=1 }^{n} {X_j^*} -n\alpha\beta \bigg) 
\end{align*}

The derivation of the \textbf{Bayesian Estimator ($f_\text{BE}$)} and its properties are deferred to Appendix \ref{apx:be}.

\begin{prop}{The \textbf{Bayesian Estimator} {(\normalfont{$ f_{\textbf{BE}}$)}} is asymptotically \textbf{unbiased} in $n$ and $m$.}
\end{prop}

\noindent Proof Sketch.
$\lim\limits_{m\to\infty}\text{Bias}(f_{\text{BA}}) = \lim\limits_{n\to\infty}\text{Bias}(f_{\text{BA}}) = 0. \hfill \blacksquare$

%\mathbb{E}\bigg[ \frac{m+ \alpha^2 n}{(m+ \alpha^2 n + \lambda_0)}\mu \bigg]$

Note that even when $n$ and $m$ are small, we can reduce the bias of the Bayesian Estimator to an arbitrarily small value by choosing very small values of the prior precision, $\lambda_0$. Thus in practice, the Bayesian estimator can be considered unbiased for small sample sizes. We want to further characterize the efficiency of the Bayesian approach as compared to the Conservative and Greedy approaches.

\begin{restatable}{prop}{bevsce}{The \normalfont{$\text{MSE}(f_{\text{BE}}) \leq \text{MSE}(f_{\text{CE}})$} iff.
$$\nicefrac{\sigma_*^2}{\sigma^2} \leq \big(\nicefrac{n}{m} \cdot \alpha^2 + 2 \big). $$
\label{prop:bevsce}}
\end{restatable}
\vspace{-1em}
\noindent Proof Sketch. Set $\text{Var}(f_{\text{BE}})\leq \text{Var}(f_{\text{CE}})$,
and the positivity of $\sigma^2$, $\sigma_*^2$, $n$  and  $m$. \hfill \hfill \hfill\hfill\hfill\hfill(Appx. \ref{apx:bevsce}) $\blacksquare$

\begin{cor*}{If $\nicefrac{\sigma_*^2}{\sigma^2} \leq 2$, \normalfont{$\text{MSE}(f_{\text{BE}}) \leq \text{MSE}(f_{\text{CE}}).$ }}
\end{cor*}

\noindent Proof. Follows immediately from Proposition \ref{prop:bevsce}. $\hfill\blacksquare$

From Proposition \ref{prop:bevsce} and its Corollary, we see that as long as the relative variance of the \textit{bad} instruments is bounded, the Bayesian estimator recovers the true $X$ more efficiently than the Conservative Estimator, independent of the actual coefficient of linear miscalibration, $\alpha$. In comparison, in situations where the \textit{bad} instruments underestimate ($|\alpha|<1$), the efficiency for the greedy approach requires the \textit{bad} instruments to have much smaller relative variances, as is evident from Propositions \ref{prop:gevsce} and \ref{prop:bevsce}. This suggests that the greedy estimator's efficiency deteriorates rapidly with smaller values of $\alpha$. In comparison, the Bayesian Estimator is robust.
 
\begin{figure*}[!htbp]
\centering
\begin{minipage}{.329\textwidth}
  \centering
  \includegraphics[width=\linewidth]{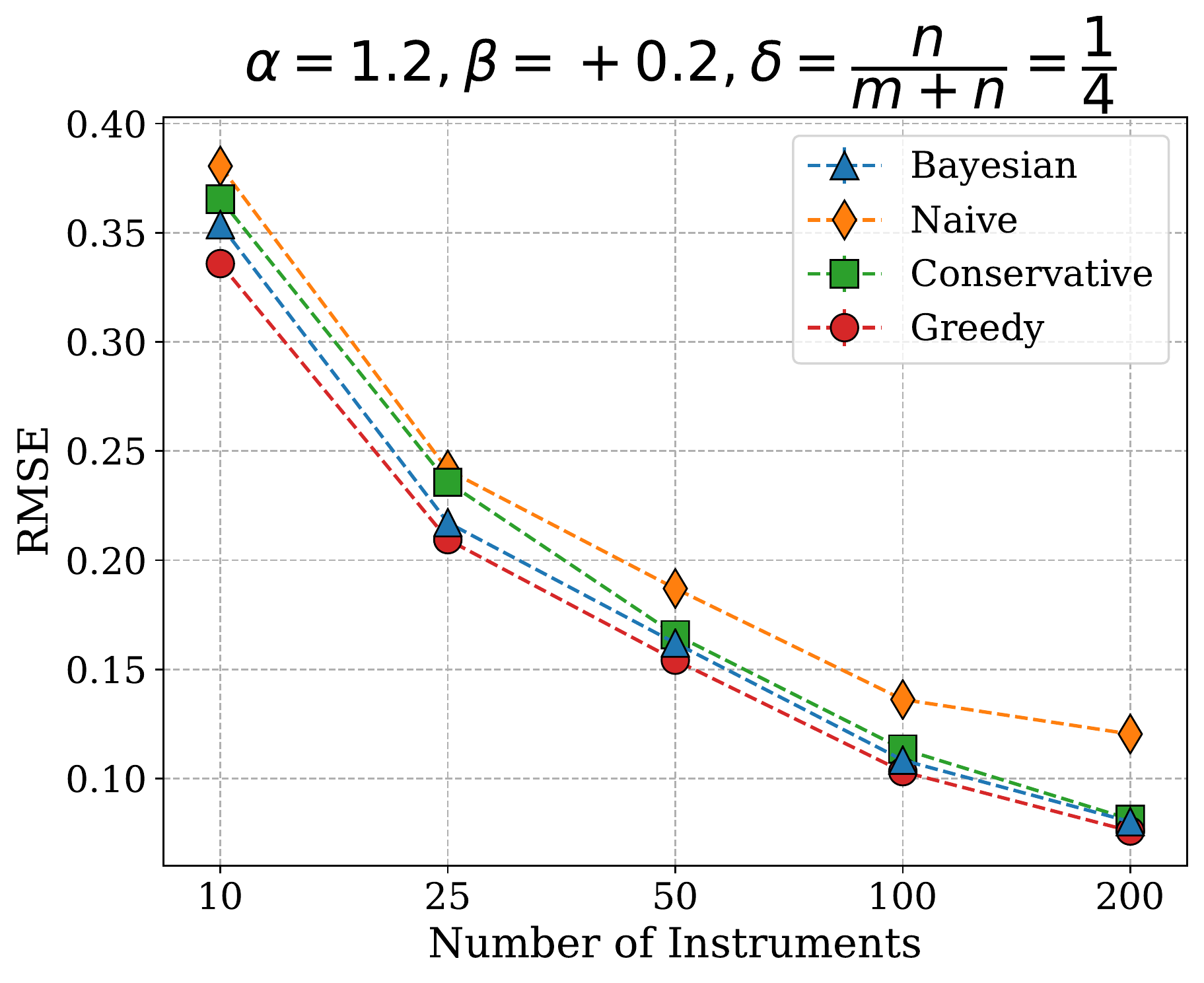}
\end{minipage}\hfill
\begin{minipage}{.329\textwidth}
  \centering
  \includegraphics[width=\linewidth]{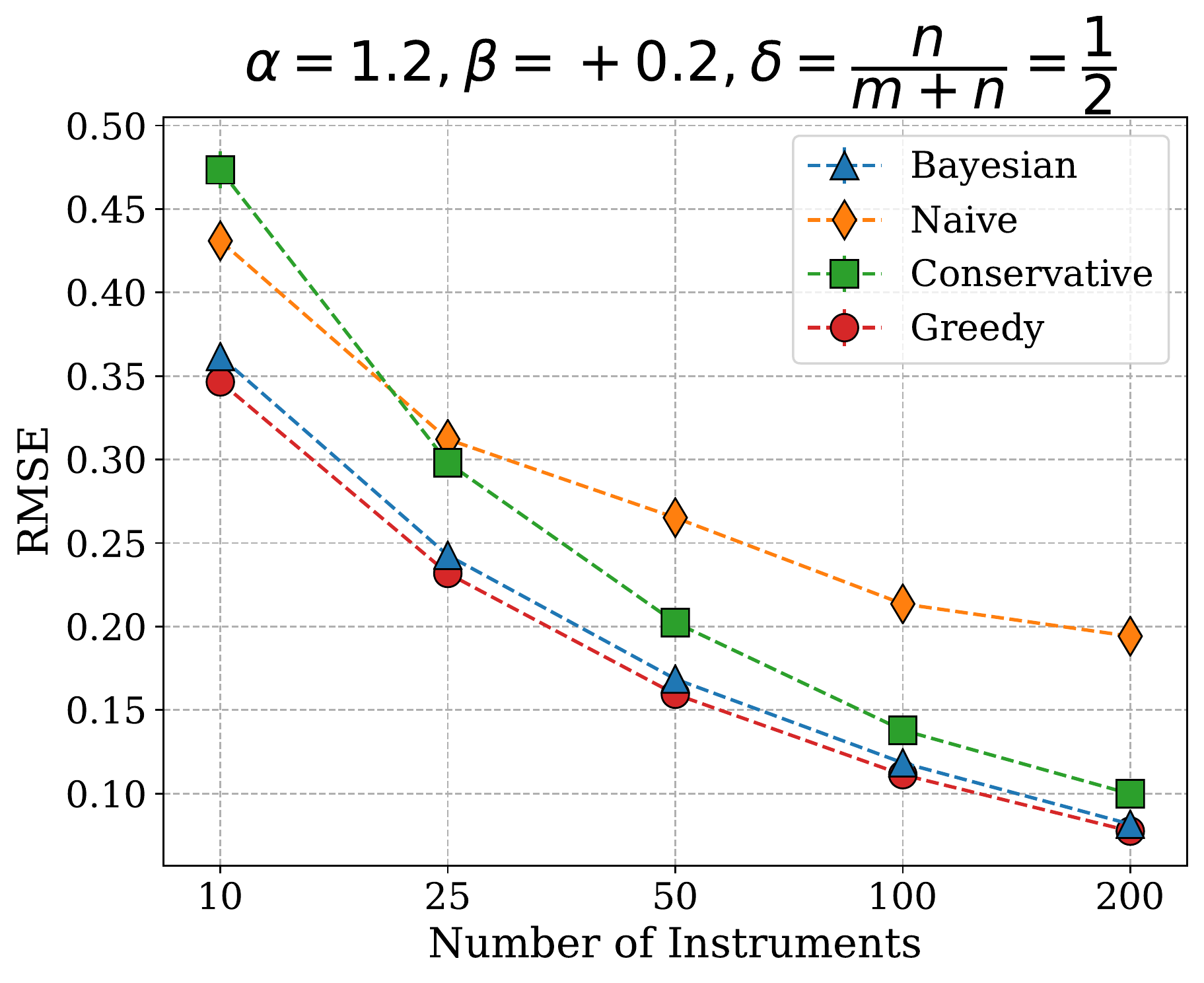}
\end{minipage}\hfill
\begin{minipage}{.329\textwidth}
  \centering
  \includegraphics[width=\linewidth]{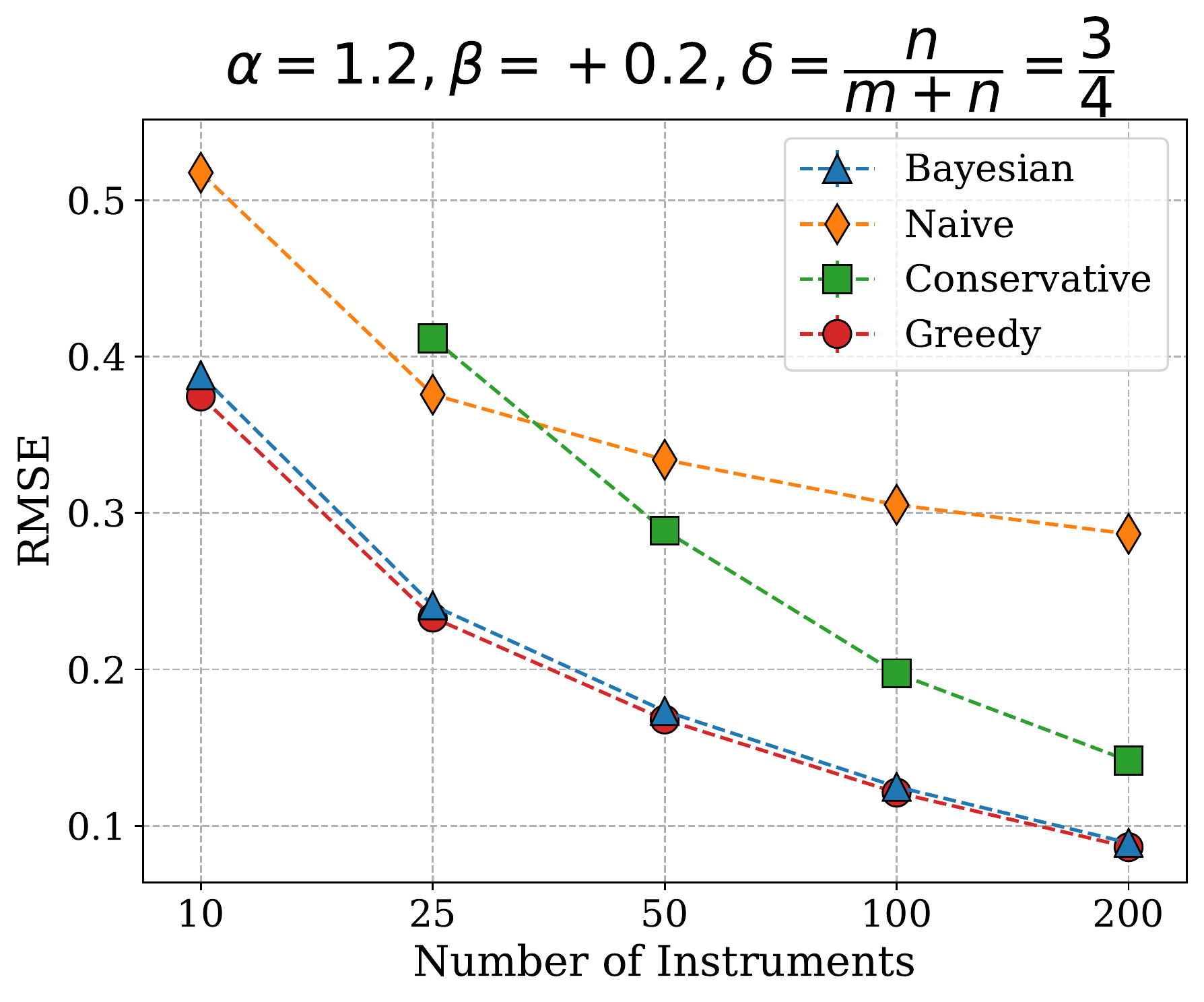}
\end{minipage}
\captionof{figure}{Mean RMSE for 1000 realizations of Synthetic Data with instruments that overestimate. }
\label{fig:mse1}
\end{figure*}

\begin{figure*}[!htbp]
\begin{minipage}{.329\textwidth}
  \centering
  \includegraphics[width=\linewidth]{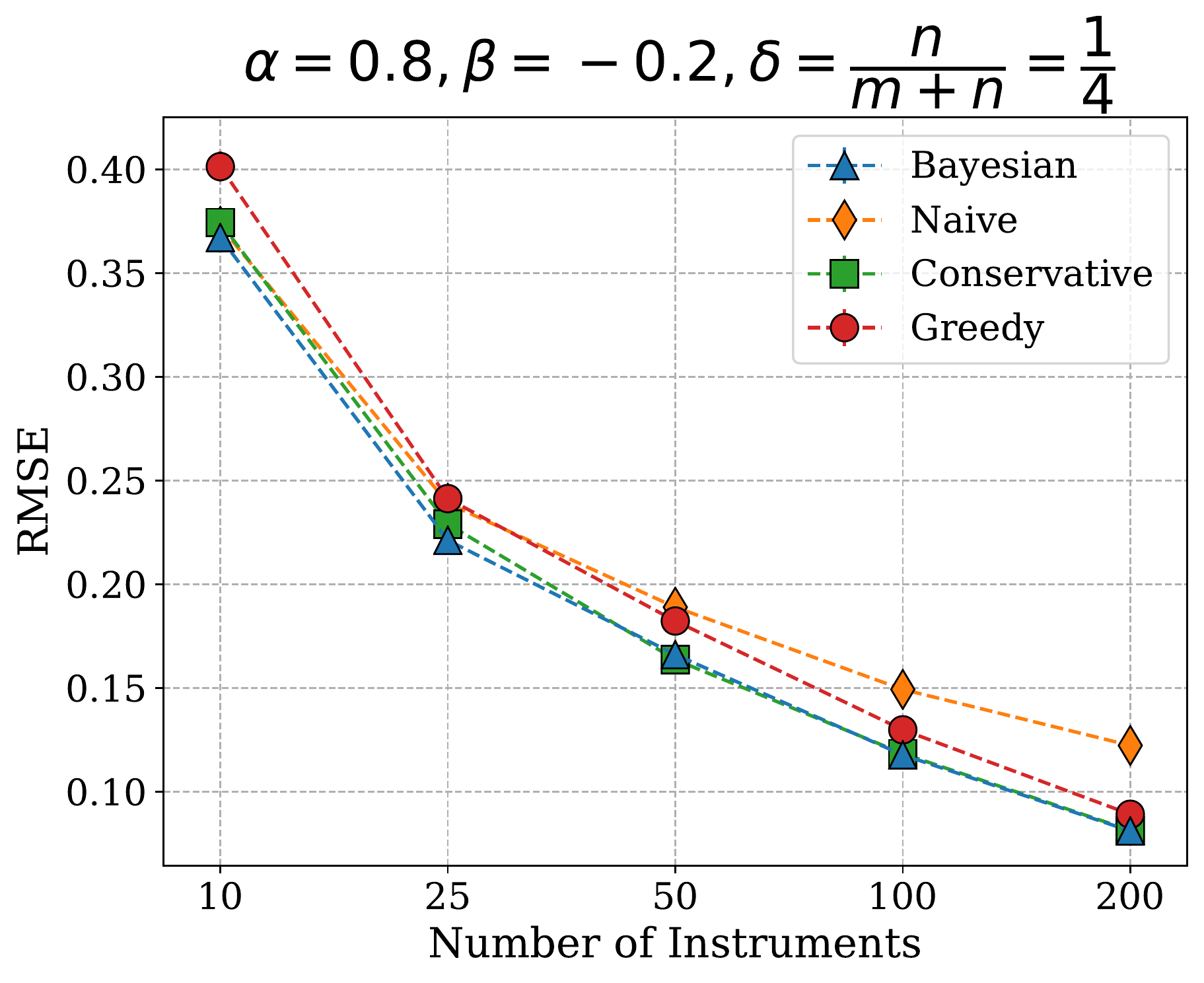}
\end{minipage}\hfill
\begin{minipage}{.329\textwidth}
  \centering
  \includegraphics[width=\linewidth]{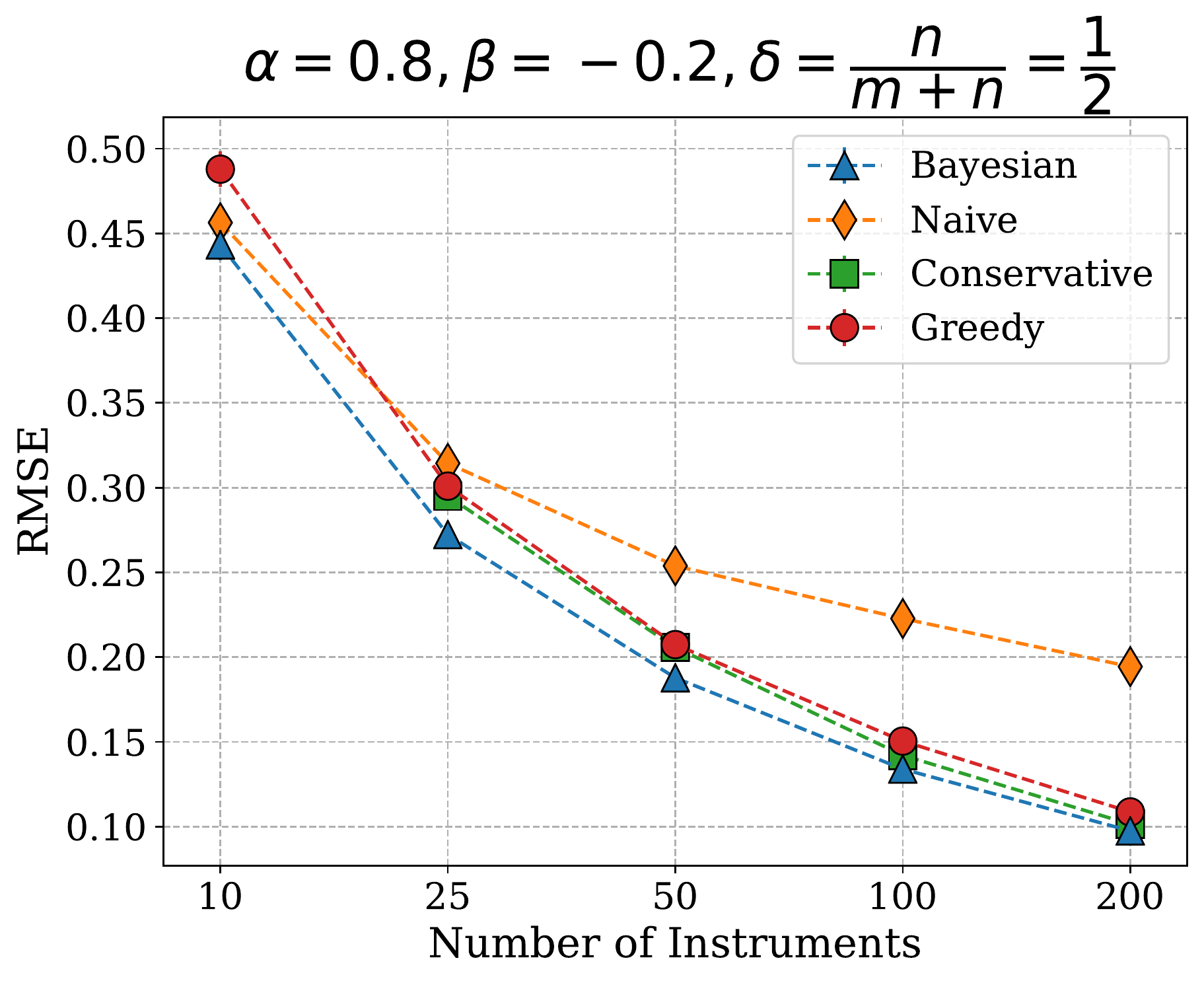}
\end{minipage}\hfill
\begin{minipage}{.329\textwidth}
  \centering
  \includegraphics[width=\linewidth]{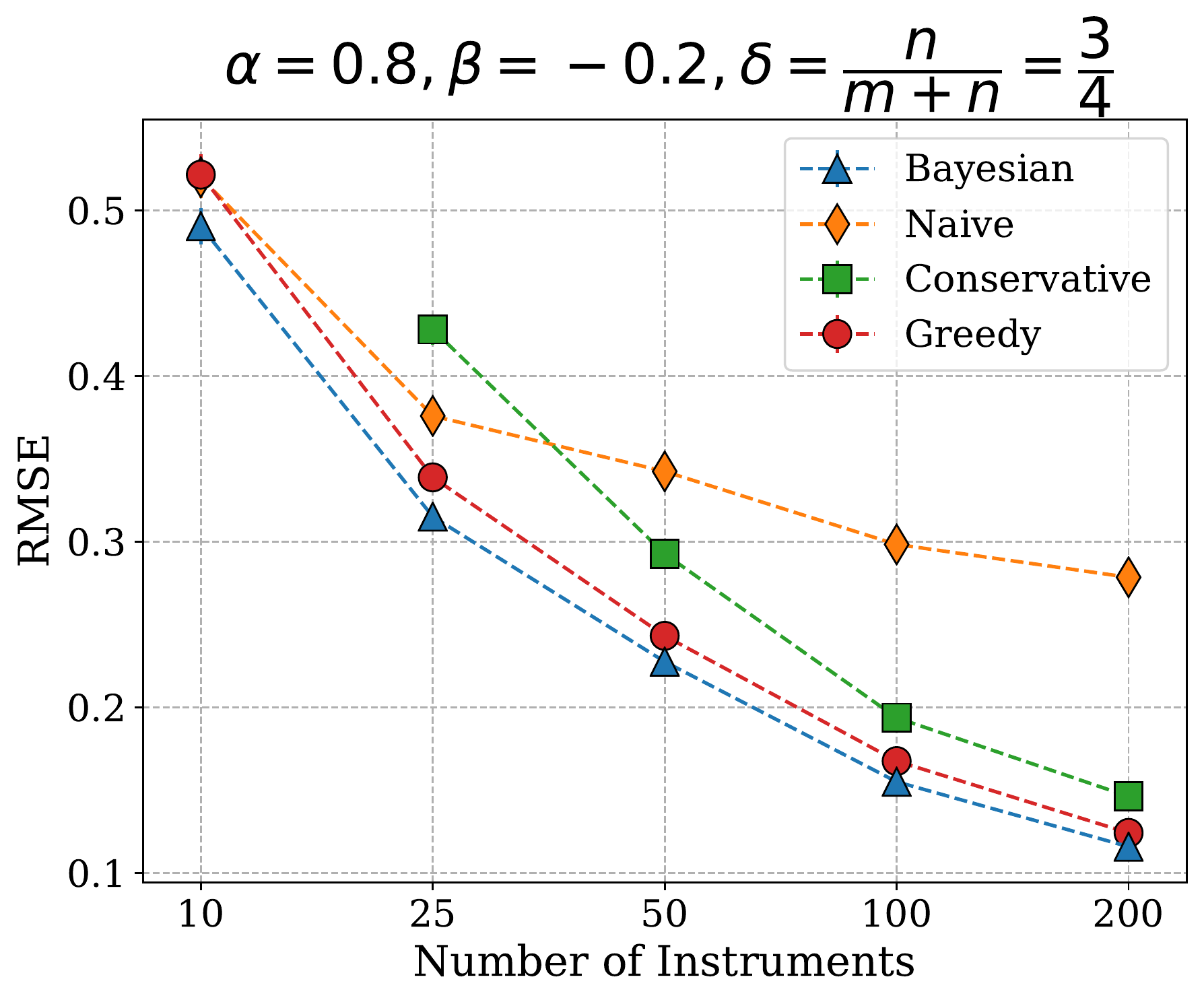}
\end{minipage}
\captionof{figure}{Mean RMSE for 1000 realizations of Synthetic Data with instruments that underestimate. }
\label{fig:mse2}
\end{figure*}

Intuitively it might seem as if the increase in performance in the previous regime comes at the cost of the Bayesian Estimator performing worse in the regime of overestimation, $|\alpha|>1$. A natural next question that arises is in regimes where $|\alpha|>1$ and $\nicefrac{\sigma_*^2}{\sigma^2} \leq 2$, how does the efficiency of the Bayesian estimator compare vis-à-vis the Greedy Estimator? Is the Greedy Estimator arbitrarily better than the Bayesian Estimator in this regime? The answer is that even in this regime, there are values of $\alpha, m $ and $n$ for which the Bayesian Estimator results in lower MSE. Since the Greedy Estimator is not strictly superior to the Bayesian Estimator, it has no real utility over the Bayesian Estimator even when instruments overestimate, as verified in Proposition \ref{prop:noutility}.

\begin{restatable}{prop}{seven}For $|\alpha| > \sqrt {\nicefrac{3}{2} + \sqrt{\frac{9}{4} + \frac{2m}{n}}} $ and  $\nicefrac{\sigma_*^2}{\sigma^2} \leq 2$,\\ \normalfont{$ \text{MSE}(f_{\text{BE}}) \leq \text{MSE}(f_{\text{GE}}) $}.
\label{prop:noutility}
\end{restatable}
Proof Sketch. The proof involves setting $\text{Var}(f_{\text{GE}})\geq \text{Var}(f_{\text{BE}})$, followed by $\nicefrac{\sigma_*^2}{\sigma^2}=k$. The resulting quadratic inequality in $\alpha^2$ is then solved for being positive. (Refer Appendix \ref{apx:bevsge})$\hfill \blacksquare$
\vspace{0.5em}
\begin{cor*}{There exists an $\alpha$ such that for  $\nicefrac{\sigma_*^2}{\sigma^2} \leq 2$, {\normalfont$\text{MSE}(f_{\text{BE}}) \leq \text{MSE}(f_{\text{GE}})$ and  $|\alpha|>1$}}.
\label{cor:noutility}
\end{cor*}
Proof. Immediately from above. $\hfill \blacksquare$

We empirically verify Propositions \ref{prop:one} - \ref{prop:noutility} using synthetic data in Section \ref{sec:simulations}.

\section{Simulations} 
\label{sec:simulations}
In this section, we experimentally verify the theoretical results from the previous Section \ref{sec:analysis} using synthetic data.

\subsection{Synthetic Data Generation}

\begin{tcolorbox}[enhanced, rounded corners, boxrule=.8pt,drop fuzzy shadow,colback=white,colframe=black,
  colbacktitle=white,coltitle=black]
\textbf{Generative Process}
\small

\hspace{1em} For $i \in |\mathcal{D}|,$ $X_i \sim  \text{Uniform}(-5,5)$

\hspace{2em} For $j \in A, Z_{ij} \sim  \text{Bernoulli}(\delta)$

\hspace{3em} If $Z_{ij} = 1$, $\hat{X} \sim  \mathcal{N}(\alpha\cdot X_i+\beta, \sigma_*^2)$

\hspace{3em} If $Z_{ij} = 0$, $\hat{X} \sim  \mathcal{N}(X_i, \sigma^2)$
\end{tcolorbox}

In order to simulate both situations in which the measuring instruments are prone to over and under estimation, we choose two different sets of $\alpha, \beta$ values. For the case in which we are trying to simulate over estimation, we set $\alpha$ as $1.2$ and $\beta$ as $+0.2$. For the case in which we are trying to simulate instruments which underestimate, we set $\alpha$ as $0.8$ and $\beta$ as $-0.2$.

In order to study the effects of the relative ratio of \textit{bad} instruments to the total number of instruments,  we create three different datasets with the ratio $\delta = \frac{n}{m+n} \in \{\frac{1}{4},\frac{1}{2}, \frac{3}{4}  \}$. Furthermore, for the experiments to be consistent with Propositions \ref{prop:bevsce}, we set $\sigma^2 = 1$ for the \textit{good} instruments and $\sigma_*^2 = 1.5$ for the \textit{bad} instruments, thus ensuring $\nicefrac{\sigma_*^2}{\sigma^2} \leq 2 $. For each realization of the data, we set the number of samples, $|\mathcal{D}|$ as $1000$, and vary the number of instruments $A$, between $\{10, 25, 50, 100, 200 \}$. Large numbers of samples help to get tighter confidence intervals, while varying the number of instruments allows one to appreciate the asymptotic performance of the estimators.

\subsection{Results}

We compare the Root-MSE values for the proposed estimators as a function of the total number of instruments in Figures \ref{fig:mse1} and \ref{fig:mse2}. Notice in Figure \ref{fig:mse1}, the Greedy estimator seems to have a marginal advantage as compared to the Bayesian estimator in the case where the \textit{bad} instruments are prone to overestimation $(|\alpha|>1)$.
However from Figure \ref{fig:mse2} in the case where the instruments are prone to underestimation ($|\alpha|<1$) the efficiency of the Bayesian Estimator is evident. In both the cases however, the other competing estimators are much worse, empirically confirming the advantages of the Bayesian approach. Also notice the performance of the Bayesian estimator is robust to increases in the proportion of the \textit{bad} instruments ($\delta$), as is evident from the efficiency of the Bayesian Estimator with larger values of $\delta$.

\section{Proposed Approach: Latent Variable \textit{Bayesian Consensus} (LVBC)} %\todo{CN: Harmonize QoI and X}
\label{sec:lvbc}

Motivated by the Bayesian Estimator proposed in the previous section, we now propose a Bayesian latent variable model for robust consensus estimation from noisy and biased or miscalibrated forecasts. 
% In this section, we propose a Bayesian latent variable model for analysts' earnings estimates and show how it can be used to infer a robust adjusted consensus earnings estimate. We use historical analyst estimates of company earnings and actual reported earnings to learn the parameters of this model. We then describe a procedure for inverse inference to generate a robust consensus estimate of future earnings from individual analysts' estimates. 

\begin{figure}[!htbp]
    \centering
    \includegraphics[width=0.65\linewidth,clip]{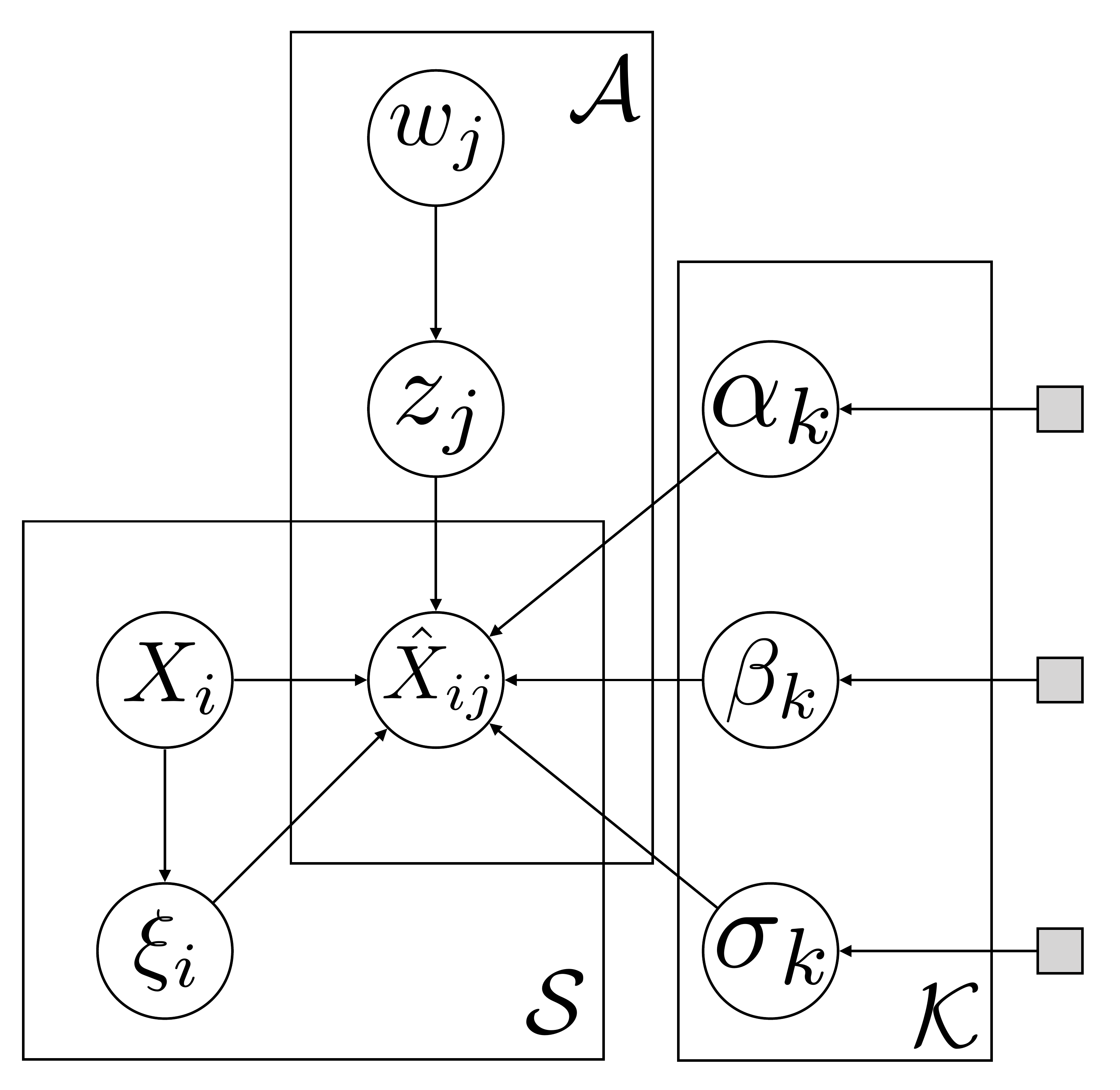}
    \captionof{figure}{The model in plate notation. $\mathcal{A}$,  $\mathcal{S}$, and $\mathcal{K}$, represent the sensors, quantities, and latent subgroups. $X_i$ is the actual change in measurement for quantity $i$ between forecast horizons and $\hat{X}_{ij}$ is sensor $j$'s estimate.}
    \label{fig:plate_model}
\end{figure} %

We assume there exists a latent subgrouping of forecasts or sensors such that within each group, we observe similar errors. We hypothesize that forecasts or sensor estimates for \emph{changes} in the \textbf{Quantity of Interest (QoI)} (the difference between the signal forecast for the next period and the true signal from the previous period) are normally distributed around a linear function of the actual resulting change with some heteroscedastic variance. Both the variance and the parameters of the linear function are conditioned on the latent subgroup a forecast or sensor belongs to. In this setting, the distribution of the \emph{forwards model}, or the observed forecast error process, can be written in closed form and parameter learning can be carried out with a gradient based method.

We propose a gradient based variational inference procedure for parameter learning and a Gibbs Sampling based MCMC algorithm for aggregating the individual estimates to generate a consensus estimate under the learned model at test time. Figure \ref{fig:plate_model} describes our model in plate notation.

\subsection{The Forwards Model} %\todo{Add the priors on the $\alpha, \beta, \sigma$}

For a particular forecast period, we use $X_i$ to represent the actual change in the QoI from the previous period, where $i$ is an index over the set of quantities being forecasted, $\mathcal{S}$, and $\xi_i$ is an indicator of whether the change in the QoI is positive or negative. Now, for each sensor indexed by $j$ in the set of sensors, $\mathcal{A}$, we draw a categorical variable $z_j$ conditioned on the parameters $w_j$ that determines which one of the $K$ latent subgroups sensor $j$ belongs to.

Finally we draw a set of parameters $w_j$ conditioned on the sensor subgroup that interacts with the true change in the QoI $X_i$ and $\xi_i$. We show the model in plate notation in Figure \ref{fig:plate_model} and give the explicit steps in the process below:
\begin{enumerate}[leftmargin=.45cm, itemsep=2mm]
    \item Draw $\{\alpha_k,\beta_k, \sigma_k\}_{k=1}^{K}$ from their priors as,\\
    \hspace*{1.75em} $ \alpha_k \sim \mathcal{N}(\widetilde{\alpha}, \nicefrac{1}{\lambda}),
    \beta_k \sim \mathcal{N}(\widetilde{\beta}, \nicefrac{1}{\lambda}),
    \sigma_k \sim \mathcal{N}(\widetilde{\sigma}, \nicefrac{1}{\lambda})$
    \item For all $j \in \mathcal{A}$, Draw $z_j  |  w_j$ as,\\
    \hspace*{.01em}$\qquad z_j| w_j \sim \text{Discrete}(\textsc{Soft-Max}(w_j))$
    \item For all $i \in \mathcal{S}$, Draw $\xi_i | X_i$ as,\\ 
    \hspace*{.1em}$\qquad \xi_i = \mathbbm{1}\{X_i>0\}$ 
    \item For all  $i \in \mathcal{S}$  and  $j \in \mathcal{A}$ ,\\ Draw $\hat{X}_{ij} | X_i, \{z_j\}_{j=1}^{\mathcal{A}}, \{ \alpha_k, \beta_k,\sigma_k \}_{k=1}^{K}$ as, \\
    \hspace*{2em}$\hat{X}_{ij}| \cdot  \sim \mathcal{N}(\alpha_{z_j}\! \! \cdot X_i + \beta_{z_j}  \;  , \; \sigma_{z_j}^{2})$
\end{enumerate}

\subsection{Parameter Learning} %\todo{explain what $\Omega$ is}

% To learn the parameters $\Theta$ of the model from observed analyst estimates and actual reported EPS data $\mathcal{D}$, we want to maximize the likelihood, {\small $P(\mathcal{D}| \Theta) = P(\{\{\hat{X}_{ij}\}_{j=1}^{|\mathcal{A}|}\}_{i=1}^{\mathcal{|S|}}, \{X_i, \xi_i\}_{i=1}^{|\mathcal{S}|}|\{\theta_j\}_{j=1}^{|\mathcal{A}|}, \{ \alpha_k, \beta_k, \sigma_k  \}_{k=1}^{K})$}, which we can rewrite as follows: 
% \begin{align*}  
% P(\mathcal{D}| \Theta)&= \prod^{|\mathcal{S}|}_{i=1} \prod_{j=1}^{|\mathcal{A}|} \sum_{k=1}^{K}  P_k(\hat{X}_{ij}|X_i, \xi_i, z_j) P(z_j|\theta_j) \quad (\text{Here, } P_k \text{ is as in Eq.  \ref{eqn:Pk}})  \\
% \log P(\mathcal{D}| \Theta) &= \sum^{|\mathcal{S}|}_{i=1} \sum_{j=1}^{|\mathcal{A}|} \log \sum_{k=1}^{K}  P_k(\hat{X}_{ij}|X_i, \xi_i, z_j) P(z_j|\theta_j)\\
%  %&= \sum^{|\mathcal{S}|}_{i=1} \sum_{j=1}^{|\mathcal{A}|} \log \mathbb{E}_{z_j|\theta_j} \big[  P_k(\hat{X}_{ij}|X_i, \xi_i, z_j) \big] \\
% &\geq \sum^{|\mathcal{S}|}_{i=1} \sum_{j=1}^{|\mathcal{A}|}  \mathbb{E}_{z_j|\theta_j} \big[ \log P_k(\hat{X}_{ij}|X_i, \xi_i, z_j) \big] \quad \text{(From Jensen's Inequality)}\\
% &\triangleq \textbf{ELBO}(\Theta; \mathcal{D})
% \end{align*}

% Here, $\textbf{ELBO}(\Theta; \mathcal{D})$ is the \emph{Evidence Lower Bound}, which is commonly used when carrying out Variational Inference with Graphical Models \citep{blei16}. In the vernacular of Variational Inference, our approximating distribution, $q(Z)$ of the True Posterior is the posterior distribution of $Z$ given $\theta$, $p(Z_i|\theta_i)$.

% \subsection{Optimization}

For learning, we maximize the \textbf{ELBO} as derived in Appendix \ref{apx:elbo} in the following explicit form.
\begin{align*}
\textbf{ELBO}(\bm{\theta}; \mathcal{D})=\\
 -\sum\limits_{i=1}^{|\mathcal{S}|} \sum\limits_{j=1}^{|\mathcal{A}|}\sum\limits_{k=1}^{K} \Tiny{\textsc{Soft-Max}}^{(k)}(w_j) \cdot \bigg( \frac{|| (\alpha_k^{(\xi_i)} \cdot \hat{X}_{ij} +\beta_k^{(\xi_i)} ) - X_i ||_{2}^{2}}{2\sigma^2_{k}} -\ln\sigma_k \bigg)\\
 - \lambda \mathbf{\Omega}(\bm{{\theta }})
\end{align*}

Here, $\mathbf{\Omega}(\bm{{\theta }}) =  \sum\limits_{k}  \big( ||\alpha - \widetilde\alpha ||_{2}^{2} + ||\beta -\widetilde\beta ||_{2}^{2} + ||\sigma -\widetilde\sigma ||_{2}^{2} \big)$ is the effect of the priors on the parameters. For purposes of identification and to ensure convergence to a good local minimum, we fix the hyper parameters corresponding to the latent group $K=1$ as $\alpha=1$ and $\beta=0$. Thus, the semantic interpretation of this latent group is that estimates from this group are accurate and unbiased.

We perform optimization using the popular first order optimizer \textsf{Adam} \citep{adam} with a learning rate of $1\times 10^{-4}$. 
For a complete description of the experiments, including the choice of hyper parameters and minibatches, please refer to Appendix \ref{apx:exp}.

We minibatch the outer two summations with a minibatch size of 5000. Early stopping is performed and optimization is terminated as soon as we overfit the validation set.

% \subsection{Parameter Initialization}

% For all subgroups $K\neq1$, we initialize $\alpha_k$ and $\beta_k$ using the coefficient of Ridge regression estimates by regressing the estimates $\{ \{\hat{X}_{ij}\}_{j=1}^{|\mathcal{A}|} \}_{i=1}^{N}$ on the corresponding set of changes in actual EPS $\{X_i \}_{i=1}^{N}$.
% We further set the initial variance of each latent group $\sigma^2_k$ to $1.0$. We observe that in practice using these initial values leads to better convergence.

\subsection{Inverse Inference} 

At test time, we want to infer a robust estimate for the change in the true quantities from the predictions $X_{ij}$ and the learned parameters $\Theta$. For quantity $i$, this is equivalent to inferring $P\left(x_i| \{ \hat{x}_{ij} \}_{j=1}^{|\mathcal{A}|}, {\Theta} \right)$. 

In our formulation, inference at test time is harder than parameter learning. This is primarily because the posterior over the latent variables is intractable. We can, however, express the conditional distributions of each variable in closed form, which allows us to use Gibbs Sampling, a Markov Chain Monte Carlo technique that allows inference by sampling from the conditional distributions, to overcome this challenge. Sampling from the full conditionals is easy for all variables except the true changes in the actuals, $X_i$.

\begin{restatable}{prop}{markovblanket}{
Under the model assumptions in Figure \ref{fig:plate_model}, the Posterior Distribution of $X_i$ conditioned on its Markov Blanket in the depicted directed acyclic graph (DAG) is given as}
\label{prop:posterior}
\end{restatable}
\begin{center}
    $  X_i | \widetilde{[\mathbf{X}]}, \{z_j\}_{j=1}^{|\mathcal{A}|} \sim  \text{\text{Multivariate-Normal}}\Big( [\mathbf{X}]  \;  , [\bm{\Sigma}] \Big)$
\end{center}
\begin{align*}
\text{where,} \quad  [\mathbf{\Sigma}] &= \big(\lambda_0 + [\bm{\alpha}]^{\top}[\bm{\sigma}^2]^{-1}[\bm{\alpha}] \big)^{-1},\\
\quad [\mathbf{X}] &= [\mathbf{\Sigma}] ([\bm{\alpha}]^{\top} [\bm{\sigma}^2]^{-1}\widetilde{[\mathbf{X}]})
\end{align*}
\begin{center}
$\text{and, } {\tiny{ \widetilde{[\mathbf{X}]} =  \left[ \begin{smallmatrix}
    \hat{X}_{i0} - \beta_{z_0}       \\
    \hat{X}_{i1} - \beta_{z_1}      \\
    \vdots \\
    \hat{X}_{i|\mathcal{A}|} - \beta_{z_{|\mathcal{A}|} }   \\
\end{smallmatrix} \right]}},
[\bm{\alpha}] =\left[ \begin{smallmatrix}
 \alpha_{z_0}\\ \alpha_{z_1}\\ \vdots \\ \alpha_{z_{|\mathcal{A}|}}  \end{smallmatrix} \right],
 [\bm{\sigma}^2] = \text{diag}
 \left[ \begin{smallmatrix}
 \sigma^2_{z_0}\\ \sigma^2_{z_1}\\ \vdots \\ \sigma^2_{z_{|\mathcal{A}|}}  \end{smallmatrix} \right]$
\end{center}
\noindent Proof. Deferred to Appendix \ref{apx:posterior}

 Proposition \ref{prop:posterior} gives the posterior distribution of $X_i$ given the sensor measurements, $\{ \hat{X}_{ij} \}_{j=1}^{|\mathcal{A}|}$, and model parameters, $\{\alpha_k, \beta_k, \sigma_k \}_{k=1}^{K}$, in a closed form to allow for sampling. Algorithm \ref{alg.mainLoop} in Appendix \ref{apx:gibbssampler}
 provides the steps in the Gibbs sampling procedure for inverse inference of $X_i$.

\section{Experiments}

We evaluate our proposed LVBC consensus model and inverse inference procedure using the datasets described below for forecasting ILI percentages and company earnings.

\begin{figure*}[!htbp]
  \centering
    \begin{minipage}{.235\textwidth}
      \includegraphics[width=\linewidth]{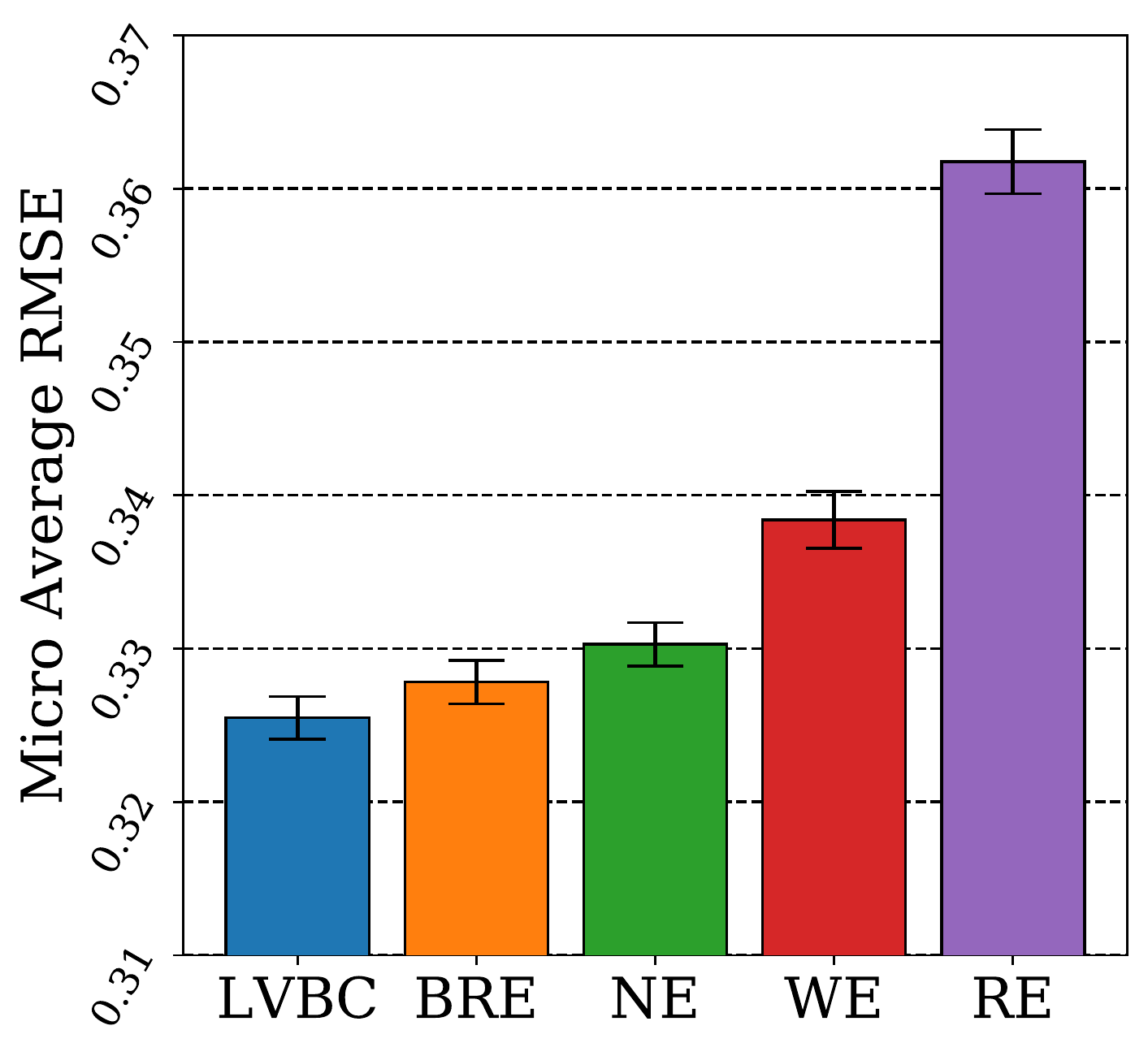}
    \end{minipage}\hspace{1.5em}
    \begin{minipage}{.235\textwidth}
      \includegraphics[width=\linewidth]{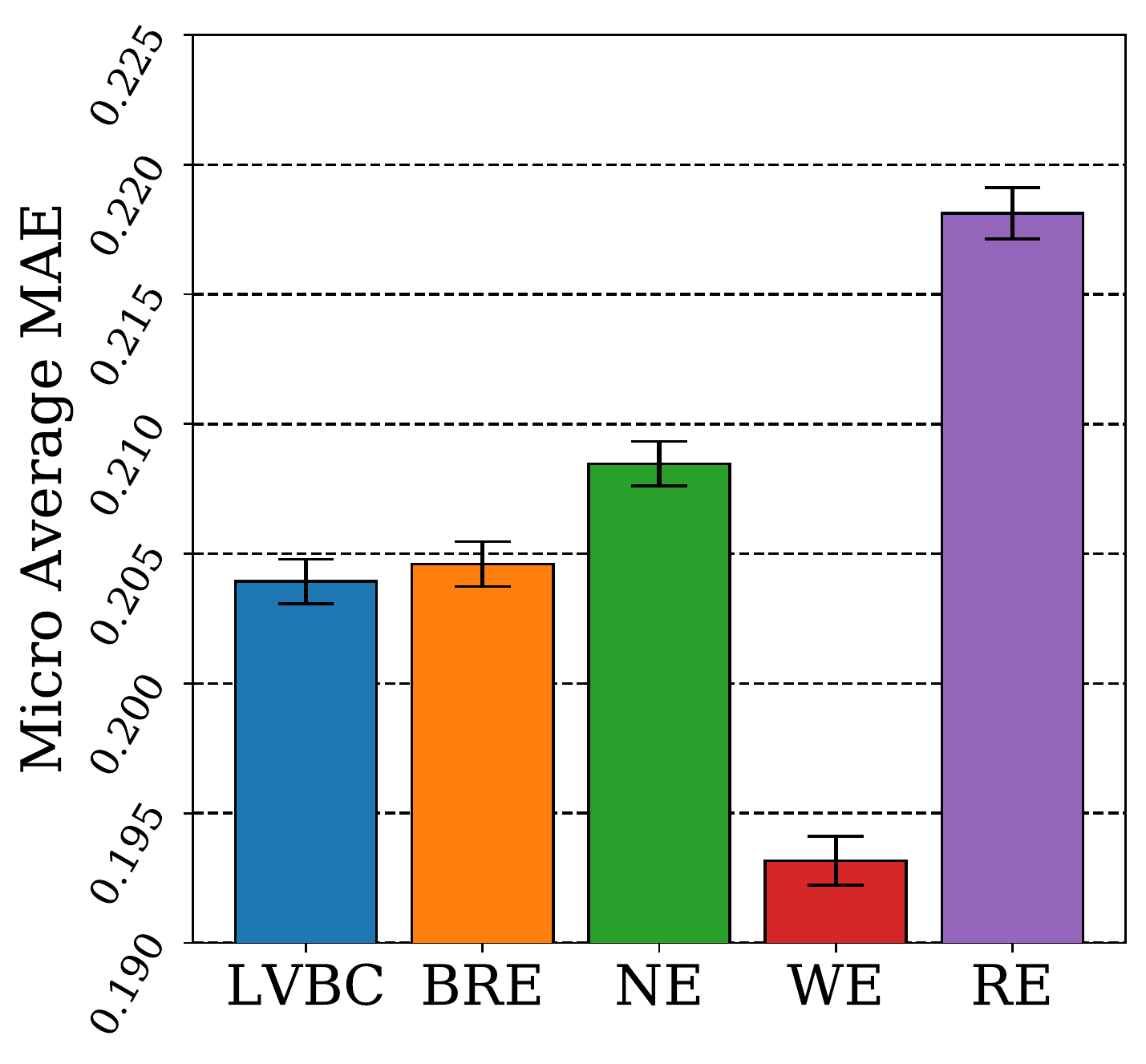}
    \end{minipage}\hspace{1.5em}
    \begin{minipage}{.235\textwidth}
      \includegraphics[width=\linewidth]{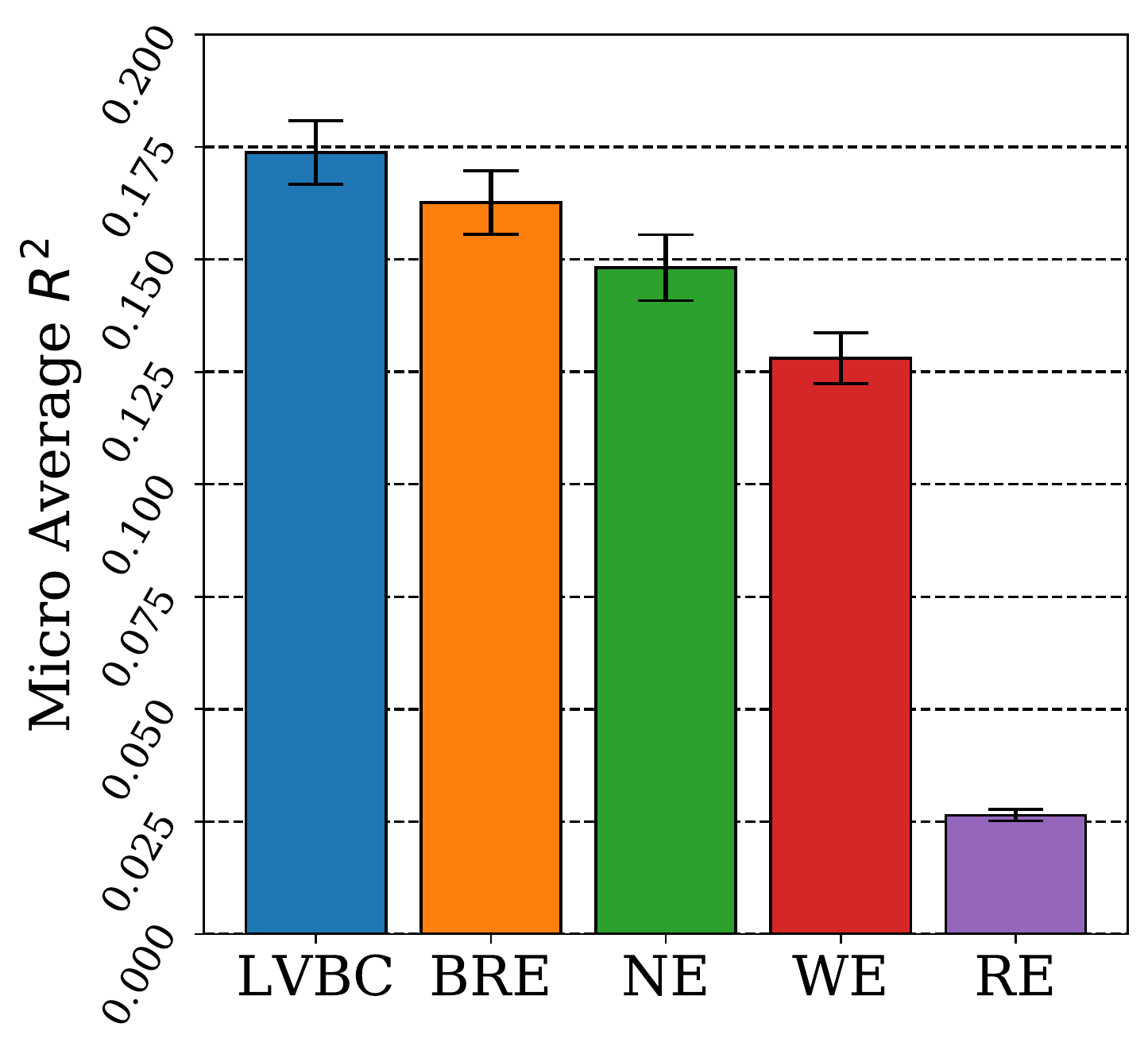}
    \end{minipage}
 \captionof{figure}{Results on the Weekly National \textbf{(ILI)} Forecasting Estimates} 
 \label{fig:ili}
\end{figure*}

\begin{figure*}[!htbp]
    \begin{minipage}{.235\textwidth}
      \includegraphics[width=\linewidth]{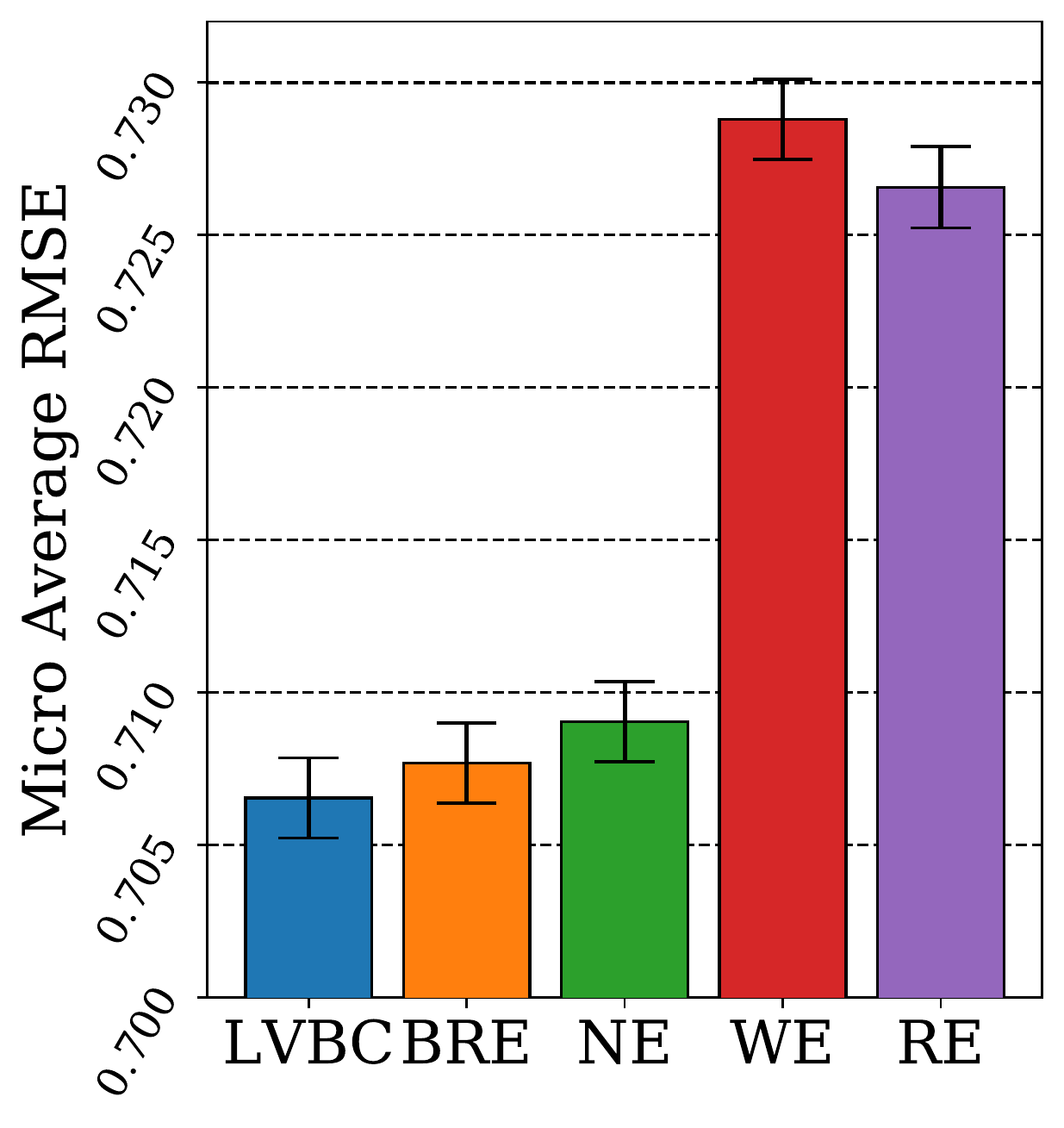}
    \end{minipage}\hspace{1.5em}
    \begin{minipage}{.235\textwidth}
      \includegraphics[width=\linewidth]{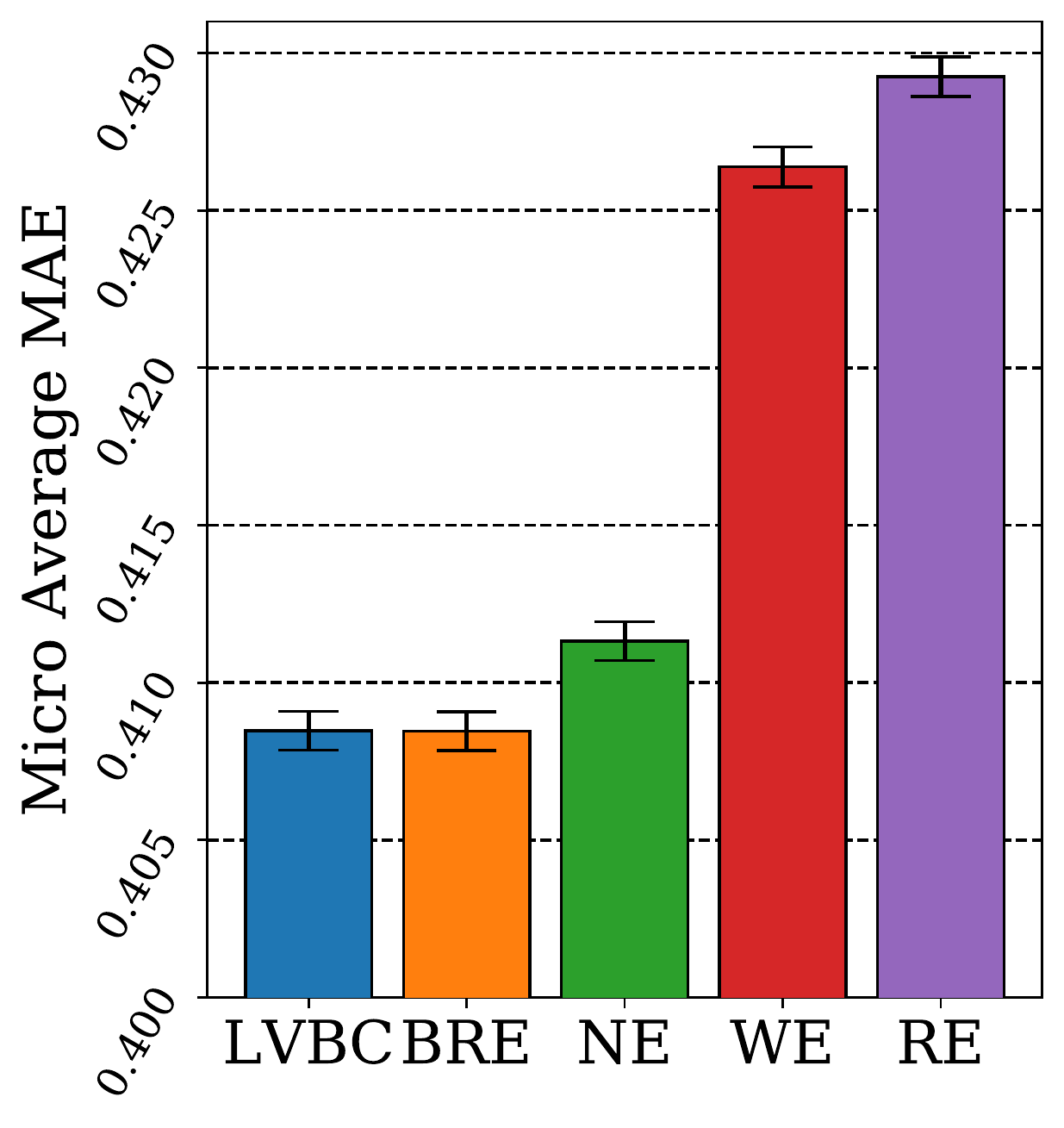}
    \end{minipage}\hspace{1.5em}
    \begin{minipage}{.235\textwidth}
      \includegraphics[width=\linewidth]{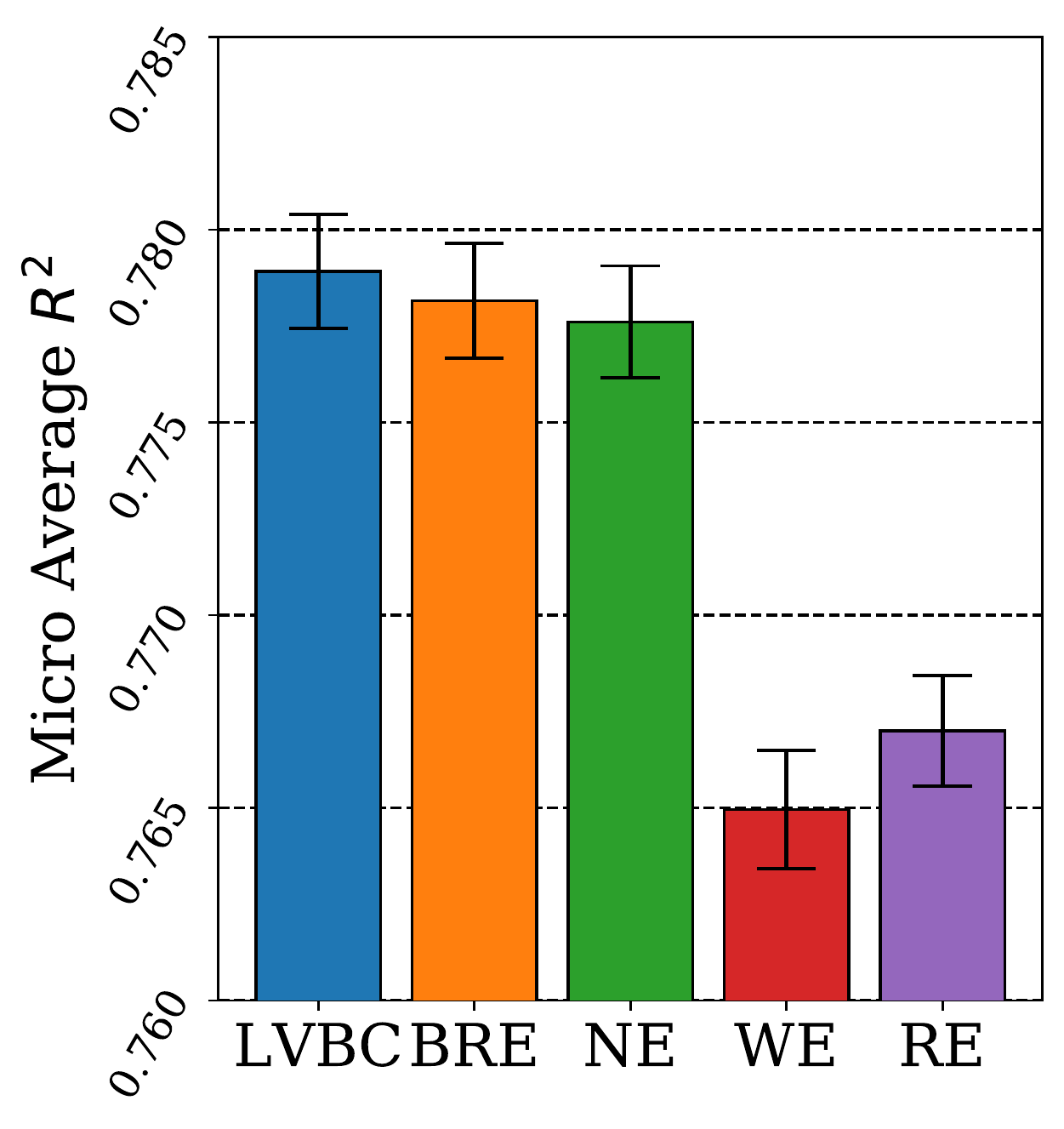}
    \end{minipage}
        \captionof{figure}{Results on the Fiscal Year 1 (\textbf{FY1}) Year Estimates} 
    \label{fig:fy1}
\end{figure*}

\begin{figure*}[!t]
    \centering
    \begin{minipage}{.235\textwidth}
      \includegraphics[width=\linewidth]{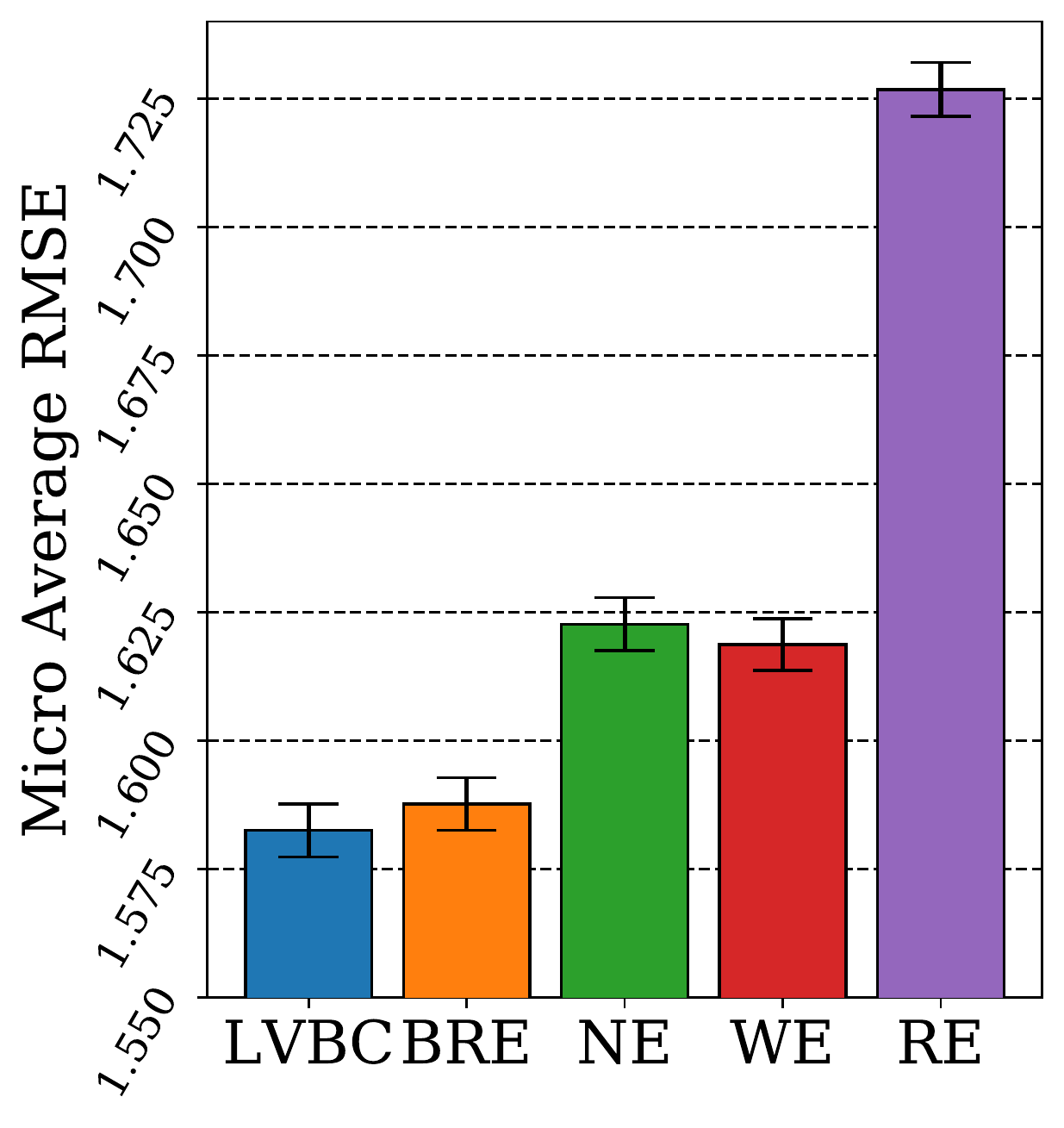}
    \end{minipage}\hspace{1.5em}
    \begin{minipage}{.235\textwidth}
      \includegraphics[width=\linewidth]{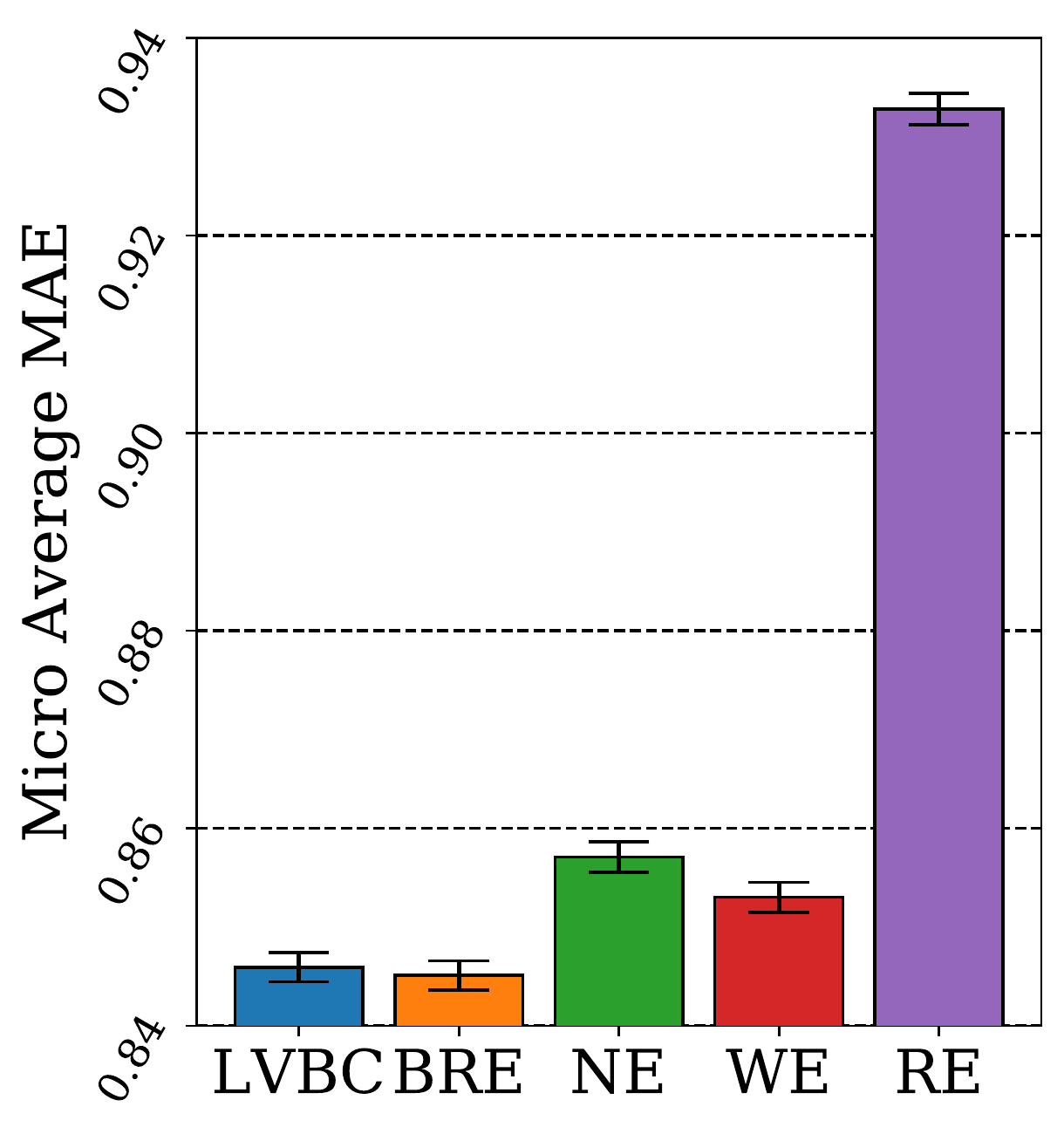}
    \end{minipage}\hspace{1.5em}
    \begin{minipage}{.235\textwidth}
      \includegraphics[width=\linewidth]{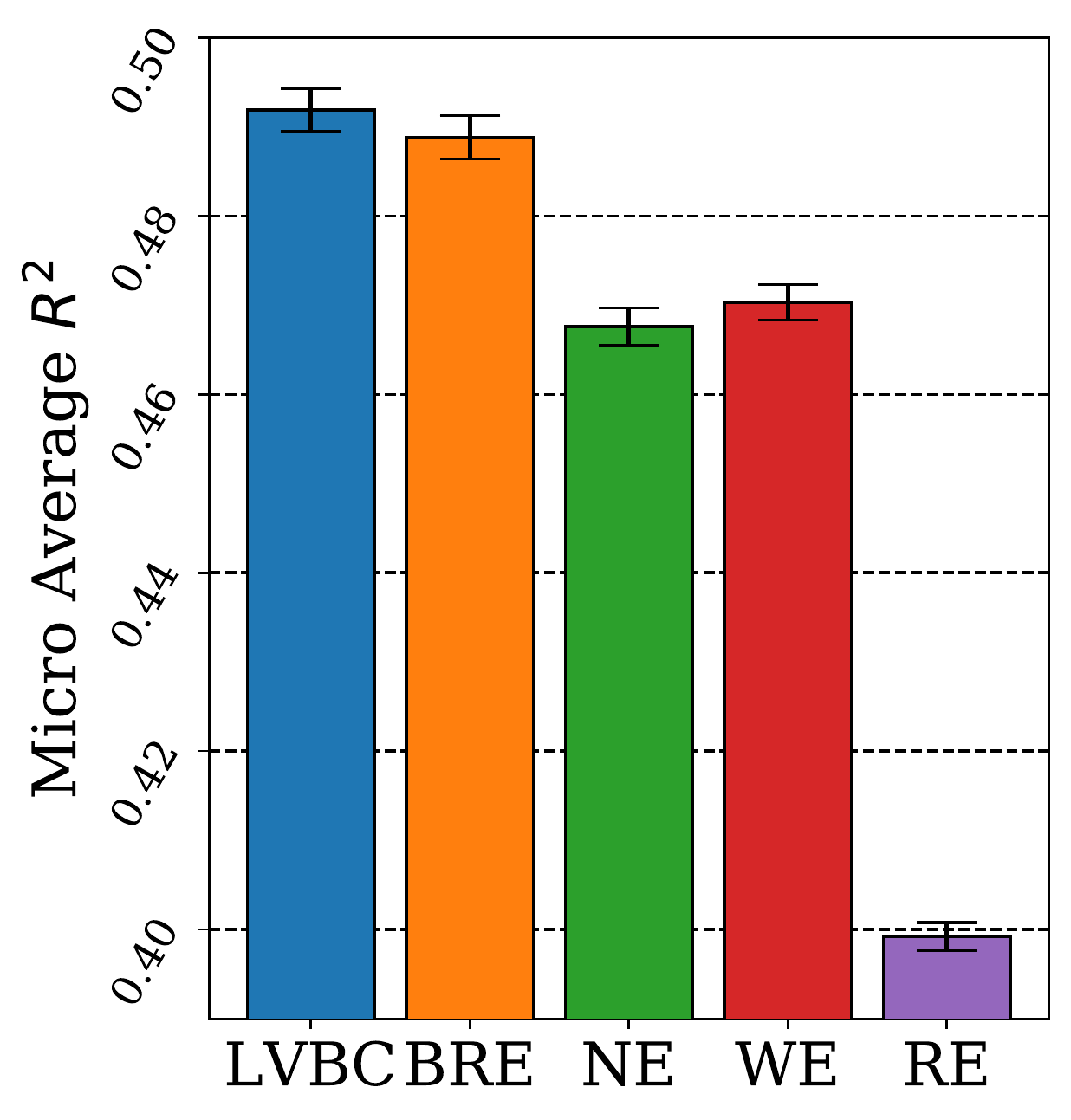}
    \end{minipage}
    \captionof{figure}{Results on the Fiscal Year 2 (\textbf{FY2}) Year Estimates} 
    \label{fig:fy2}
\end{figure*}

\subsection{Datasets}

\noindent \textbf{CMU-Delphi Flu Sensors Dataset:} This dataset consists of US National Weekly ILI percentages from $1^\text{st}$ January, 2015 through $31^\text{st}$ December, 2019 along with the forecasts made by a variety of sensors. We use weekly forecasts and actuals during $2018$ for validation and test on forecasts for all weeks of 2019. The CMU Delphi Flu Dataset was aggregated from the Delphi API\footnote{\url{https://cmu-delphi.github.io/delphi-epidata/api/}}. We considered the \textsf{SAR3, EPIC} and \textsf{ARCH} sensor forecasts for weekly \textbf{ILI} from ten of the largest states (by population) in the US, CA, TX, FL, NY, PA, IL, OH, GA and NC. Note that only \textsf{SAR3, EPIC} and \textsf{ARCH} were available publicly and used in \cite{jahja:2019}. For a full description of the sensor meanings and datasources they rely on we defer the reader to \cite{delphiapi}. The prediction task is then defined to be to use these sensor forecasts to forecast US national weekly \textbf{ILI} percentages, which were also aggregated from the Delphi API. 

\begin{table}[!htbp]
    \centering
    \begin{tabular}{|l|r|}\toprule \midrule
         Total No. of Companies & $200$ \\
         Total No. of Analysts&$7,999$\\
         Avg. No. of Analysts / Company&142.28\\ 
         Std. Dev. of Analysts / Company&28.52\\ 
         \midrule \bottomrule
    \end{tabular}
    \caption{Summary statistics of the \textbf{I/B/E/S} data subset.}
    \label{tab:statitics}
\end{table}

\noindent \textbf{Thompson Reuters Institutional Brokers Estimate System (I/B/E/S) Dataset:}\footnote{\small \url{https://www.refinitiv.com/en/financial-data/company-data/institutional-brokers-estimate-system-ibes}} This dataset consists of earnings forecasts published by analysts at major financial institutions for companies in their coverage universe along with the actual earnings reported by those companies. Multiple forecast horizons and quantities are included. We focus on forecasting \emph{Earnings per Share (EPS)}, a measure of net profit, as this is most widely used quantity. EPS is the ratio between a company's net income after subtracting preferred dividends and the number of outstanding common shares of the company's stock. For our experiments, we limit our analysis to the top 200 companies covered by the most analysts and consider two forecast horizons: forecasts for the \textbf{Next Fiscal Year (FY1)}, or forecasts for the EPS reported by companies on their next scheduled annual reporting date, and \textbf{Second Fiscal Year (FY2)}, or forecasts for the following annual reporting date. Some analysts publish multiple forecast revisions during these periods. We only consider revisions made at least 6 months (12 months) before the annual reporting date for FY1 (FY2). We use data from January $1^\text{st}$, 2000 to January $1^\text{st}$, 2012 for training, data from January $1^\text{st}$, 2012 to January $1^\text{st}$, 2014 for validation and data from January $1^\text{st}$, 2014 to January $1^\text{st}$, 2019 for testing. As compared to the previous dataset, the task involving forecasting \textbf{I/B/E/S} Dataset is subsantially harder and involves 200 companies and multiple revisions from multiple analysts. Note that not all analysts follow all the stocks, and most analysts only follow a few stocks with from industrial sector they have most experience in. Table \ref{tab:statitics} gives some statistics about the distribution of the analysts vis-à-vis the companies in our dataset.

\subsection{Evaluation}
We first learn parameters from historical sensors or earnings estimates and the actual weekly \textbf{ILI} percentages and reported earnings and then carry out inverse inference to predict changes in the actuals $X_i$ from test data forecasts $X_{ij}$. We compare these consensus forecasts made using our \textbf{LVBC} approach to the simple uniform consensus forecast and other approaches we describe below. Since \textbf{LVBC} is sensitive to local minima during learning, we perform 10 different random restarts during training and perform posterior inference over the validation set. We use the parameters that result in the best performance in terms of Root Mean Squared Error (\textbf{RMSE}) on the validation set. 

We consider the difference between the actual reported change in the QoI and the forecasted change using our method and each of the reference baselines described below. We report the \textbf{Root Mean Squared Error (RMSE)}, \textbf{Mean absolute Error (MAE)} and \textbf{Coefficient of Determination (\textit{R}$^\mathbf{2}$)} for the consensus \textbf{ILI} forecast in Figure \ref{fig:ili}. We report the same quantities, micro-averaged across all companies in the \textbf{I/B/E/S} dataset, for the consensus earnings forecasts for \textbf{FY1} in Figure \ref{fig:fy1} and \textbf{FY2} in Figure \ref{fig:fy2}. Error bars represent 95\% confidence intervals, generated by bootstrapping the inferred results for the test data 1000 times. To ensure the \textbf{RMSE} numbers are not a result of overfitting a few individual companies and our \textbf{LVBC} estimator generalizes to the majority of companies in the I/B/E/S dataset, we also report the macro-averaged \textbf{RMSE}, \textbf{MAE} and \textit{R}$^\mathbf{2}$ in Appendix \ref{apx:results}.

% \RT{It migth make things more clear if the Results are organized into one section and data, baselins, results are separate subsections}

\subsection{Reference Baselines}

\noindent \textbf{Naive Estimator (NE)}: The estimate of $X_i$ is made using the simple uniform consensus estimator, or average of all of the sensor readings $\{\hat{X}_{ij} \}_{j=1}^{|\mathcal{A}|}$\\ 
\centerline{Hence, $\hat{X}_i = \frac{1}{|\mathcal{A}|} \sum_{j=1}^{|\mathcal{A}|} \hat{X}_{ij}$.}

\noindent \textbf{Weighted Estimator (WE)}: Instead of averaging over all the sensors, $\{\hat{X}_{ij} \}_{j=1}^{|\mathcal{A}|}$ naively, we perform a weighted averaging such that $w_j \propto \nicefrac{1}{\widehat{\text{MSE}}(j)}$, i.e. the weight given to a sensor is inversely proportional to the sensor's historical forecast accuracy.\\
\centerline{Hence, $\hat{X}_i = \frac{1}{|\mathcal{A}|} \sum_{j=1}^{|\mathcal{A}|} w_{j} \cdot \hat{X}_{ij}$}

\noindent \textbf{Regression Estimator (RE)}: We regress the set of true values, $\{X_i\}_{i=1}^{N}$ against the corresponding forecasts, $\{\hat{X}_j\}_{j=1}^{|\mathcal{A}|}$. At test time, we perform the learnt regression on $\{\hat{X}_j\}_{j=1}^{|\mathcal{A}|}$ to get adjusted estimates for $X$. The final estimate is the average of the adjusted estimates.\\ 
\centerline{Hence, $\hat{X}_i = \frac{1}{|\mathcal{A}|} \sum_{j=1}^{|\mathcal{A}|} f( \hat{X}_{ij}, \theta )$,}\\ $ $ \hfill where $f (.)$ is the learnt regression function.\\
\noindent  We consider two different regression functions, a parametric ridge regression \textbf{RA-Ridge} and a non-parametric regression consisting of an Random Forest of Decision Trees \textbf{RA-Ensemble}. We include RA-Ridge in the plots and include both in the tables in Appendix \ref{apx:exp}.

\noindent \textbf{Bayesian Regression Estimator (BRE) :} Instead of regressing the actual $X_{i}$ on the forecasts, $\{\hat{X}_{ij} \}_{j=1}^{|\mathcal{A}|}$, we first learn a regression of $\{\hat{X}_{ij} \}_{j=1}^{|\mathcal{A}|}$ on the actuals $X_i $ with a linear link function $f(\beta^\top {X_i})$. At test time we condition on $\beta$ and place a weak conjugate prior on $X_i$. The final adjusted estimate of $X_i$ is then recovered as the expectation of $X_i$ under the posterior conditioned on $\{\hat{X}_{ij} \}_{j=1}^{|\mathcal{A}|}$ and the regression parameters $\theta$.\\ 
\centerline{Hence, $\hat{X_i} = \mathbb{E}_{X_i \sim p(.|\{\hat{X}_{ij} \}_{j=1}^{|\mathcal{A}|}, \theta)} [X_i]$.}
Note that Bayesian Regression Estimator is equivalent to \textbf{LVBC}  with $K=1$.

\section{Results}

The results in Figures \ref{fig:ili}, \ref{fig:fy1} and \ref{fig:fy2} show that \textbf{LVBC} is the top performing consensus estimator in nearly all cases (across metrics and forecasting challenges) demonstrating its effectiveness as an improved consensus model. In particular, we note that in the company earnings forecasting challenge, individual analyst estimates tend to err more as the forecast horizon is increased, as evidenced by higher \textbf{FY2} forecasting errors. We also see a greater performance increase in \textbf{FY2} forecasting when using \textbf{LVBC} as compared to other methods, indicating \textbf{LVBC} is a more robust consensus estimator in regimes with significant bias and error in individual forecasts. Although Weighted Averaging (WE) reduces errors in the \textbf{FY2} consensus estimates, this benefit is not significant given the large confidence intervals around the results. 

Interestingly, we observed that the Regression Estimator (\textbf{RE}) reduced the consensus error by a large margin on the training dataset, but had worse performance on the test set. This was true for both Parametric Ridge Regression and Non-Parametric Random Forest Regression, suggesting these models have a large tendency to overfit. Furthermore, the Bayesian Regression Estimator (\textbf{BRE}) has performance similar, but worse than \textbf{LVBC}. We hypothesize that this is because \textbf{BRE} does not allow for the flexibility of discovering sensors that are \textit{unbiased}. 
% \begin{minipage}{\textwidth}
% \centering
% \begin{minipage}{.2\textwidth}
%   \centering
%   \includegraphics[width=\linewidth]{plots/results/nat/nat_rmse.pdf}
% \end{minipage}\hspace{2em}
% \begin{minipage}{.2\textwidth}
%   \centering
%   \includegraphics[width=\linewidth]{plots/results/nat/nat_mae.pdf}
% \end{minipage}\hspace{2em}
% \begin{minipage}{.2\textwidth}
%   \centering
%   \includegraphics[width=\linewidth]{plots/results/nat/nat_r2.pdf}
% \end{minipage}
%     \captionof{figure}{Results on the Weekly National \textbf{(ILI)} Forecasting Estimates} 
% \label{fig:ili}
% \end{minipage}

\section{Conclusion and Future Work} 
We proposed a new approach for generating consensus forecasts from noisy and biased or miscalibrated individual forecasts and measurements and provided a theoretical analysis, which we confirmed experimentally, which shows that our proposed Bayesian Consensus estimator is unbiased and asymptotically more efficient than other consensus estimators. We applied our approach to two forecasting challenges: ILI forecasting and earnings forecasting. We found that our approach leads to more accurate forecasts than the simple consensus estimate and other baseline approaches. %While we focused on the problem of generating robust consensus earnings estimates, we note that the proposed model is applicable to any other problem where we have a quantity that is predicted or measured by multiple instruments or individuals, which may be subject to error of bias or miscalibration. %There are many applications in finance, economics, government and political science where this model could be applied such as GDP and unemployment forecasting and predicting the outcomes of elections. It could also be applied to some physical systems which rely on measurements from sensors and other devices which might be miscalibrated. 

There are some limitations in the current approach. First, we only account for linear miscalibrations. In real world forecasting scenarios, especially with human instruments, the miscalibration function could be highly nonlinear. Even in the simple case of polynomial functions, recovering the true outcome is challenging. We hypothesize that future work can involve using black box function approximators to learn the miscalibration function. Although this would make inference harder, we believe it might improve performance when large amounts of training data are available.

% Furthermore, our approach is sensitive to local optima since we relied on gradient based methods for learning. It remains to be seen that if sampling based method at learning time would alleviate this sensitivity of our approach.

% \subsubsection*{Acknowledgements}
% The authors would like to thank all the anonymous reviewers in advance for their comments and feedback.

\begin{acks}
We thank the anonymous reviewers for their comments on this manuscript. We would also like to thank all members of the J.P. Morgan AI Research team for helpful discussions and feedback.
\end{acks}

\section*{Disclaimer}

This paper was prepared for informational purposes by the Artificial Intelligence Research group of JPMorgan Chase \& Co and its affiliates (“J.P. Morgan”), and is not a product of the Research Department of J.P. Morgan.  J.P. Morgan makes no representation and warranty whatsoever and disclaims all liability, for the completeness, accuracy or reliability of the information contained herein.  This document is not intended as investment research or investment advice, or a recommendation, offer or solicitation for the purchase or sale of any security, financial instrument, financial product or service, or to be used in any way for evaluating the merits of participating in any transaction, and shall not constitute a solicitation under any jurisdiction or to any person, if such solicitation under such jurisdiction or to such person would be unlawful.   

\bibliographystyle{ACM-Reference-Format}
\bibliography{main}


\appendix
\onecolumn

\newpage
\section{Results in Section \ref{sec:analysis} (Analysis) }

\subsection{Proof of Proposition \ref{prop:one}}
\label{apx:one}

\one*
\vspace{-.75em}
\noindent Proof. It is trivial to see that,\vspace{-.75em}
    \begin{align*}
    \text{Bias}(f_\text{NE}) &=  \mathbb{E}[{f_{\text{NE}}(\mathcal{D})}]-\mu\\
    &=\frac{n}{(m+n)} \big[ (\alpha-1)\mu + \beta \big], \\
    \text{and, } \text{Var}(f_{\text{NE}}) &= \Big( \frac{1}{m+n} \Big)^2 \big[ m\sigma^2 +n \sigma_*^2  \big]
    \end{align*}
Now, $\lim\limits_{n\to\infty} \text{Bias}(f_\text{NE}) \neq 0 . \hfill  \blacksquare   $

\subsection{Proof of Proposition \ref{prop:two}}
\label{apx:two}

\two*
%\vspace{-1em}
\noindent Proof. It is trivial to see that, $\text{Bias}(f_{\text{CE}} ) =  \mathbb{E}[f_{\text{CE} }] - \mu = 0$ and  $ \text{Var}[f_{\text{CE}}]  =\frac{\sigma^2}{m}.  \hfill  \blacksquare $
    
\subsection{Proof of Proposition \ref{prop:three}}
\label{apx:three}

\three*
%\vspace{-1em}
\noindent Proof.
$$ f_{\text{GE}}(\mathcal{D}) = \frac{  \sum\limits_{i=1}^{m} \hat{X}_i  + \frac{1}{\alpha}\left( \sum\limits_{j=1}^{n} {X}^*_i-n\beta\right)}{m+n} $$ 

Proof. It is trivial to see that, $\text{Bias}(f_{\text{GE}} ) =  \mathbb{E}[f_{\text{GE} }] - \mu = 0 . \hfill \blacksquare$
    
\subsection{Proof of Proposition \ref{prop:gevsce}}
\label{apx:gevsce}

\gevsce*
%\vspace{-1em}
\noindent Proof. The Greedy $(f_{\text{GE})}$ and the Conservative ($f_{\text{CE}}$) estimators are both unbiased. Thus, comparing their Mean Squared Errors, reduces to comparing their Variances.
Now, 
\begin{align*}
f_{\text{GE}}(\mathcal{D}) &= \frac{  \sum\limits_{i=1}^{m} \hat{X}_i  + \frac{1}{\alpha}( \sum\limits_{j=1}^{n} {X}^*_i-n\beta)}{m+n} \text{ and } \text{Var}[f_{\text{GE}}] = \frac{  m \sigma^2  + \nicefrac{n\sigma_*^2}{\alpha^2}}{ (m+n)^{2}} 
\end{align*}
\begin{alignat*}{3}
&\text{Now, } \text{MSE}(f_{\text{GE}})&< \text{MSE}(f_{\text{CE}}) &\Longleftrightarrow \text{Var}[f_{\text{GE}}] <\text{Var}[f_{\text{CE}}]\\
&\Longleftrightarrow & \frac{  m \sigma^2  + \nicefrac{n\sigma_*^2}{\alpha^2}}{ (m+n)^{2}} &< \frac{   \sigma^2}{ m} \\
&\Longleftrightarrow & \frac{  m  + \nicefrac{n\sigma_*^2}{\alpha^2 \sigma^2}}{ (m+n)^{2}} &<\frac{ 1}{ m}\\
&\Longleftrightarrow & \frac{  m  + \frac{n\sigma_*^2}{\alpha^2 \sigma^2}}{ (m+n)^{2}} &<\frac{ 1}{ m}\\
&\Longleftrightarrow &   m^2  + \frac{mn\sigma_*^2}{\alpha ^2  \sigma^2}&<{ (m+n)^{2}}\\
&\Longleftrightarrow &   m^2  + \frac{mn\sigma_*^2}{\alpha ^2  \sigma^2}&< m^2+2mn + n^2\\
&\Longleftrightarrow &   \frac{m\sigma_*^2}{\alpha ^2  \sigma^2}&<2m + n\\
&\Longleftrightarrow &   \frac{\sigma_*^2}{\sigma^2}&<  \bigg(2 + \frac{n}{m} \bigg)\alpha^2.
\end{alignat*} 
$\hfill \blacksquare$
\newpage
\subsection{The Bayesian Estimator ($f_{\text{BE}})$ }
\label{apx:be}

In this section we describe the assumptions we make on the data generating process and derive the Bayesian Estimator under the aforementioned assumptions. The key idea here is to place a wek normal prior on $X$, the quantity we seek to recover. We further study the Bias and Variance properties of the recovered estimator.

\subsubsection{Derivation}

\textbf{Assumptions}:

1. Sample $X$ from its prior. \\
\indent$\qquad X \sim \mathcal{N}(0, \nicefrac{1}{\lambda_0})$. 

2. Conditioned on Z and assuming homoscedasticity, we sample the measurement, $\hat{X}$\\
\indent$\qquad \hat{X}_i|(Z_i=0) \sim \mathcal{N}(\mu, \mathbf{1})$.\\
\indent$\qquad {X}_i^*|(Z_i=1) \sim \mathcal{N}(\alpha\cdot\mu+\beta, \mathbf{1})$. 

Now using the fact that the measurements are independent, we can rewrite the generated samples as being drawn from a multivariate normal. 

$$\widehat{[\mathbf{X}]} | [\mathbf{Z}] \sim \text{Multivariate-Normal} ( [\mathbf{X}]  , \mathbf{1}) $$
$$\text{where,  } \widehat{[\mathbf{X}]} = \left[ \begin{smallmatrix}
    \hat{X}_{1} \\
    \vdots \\
    {X}^*_{1} \\
    \vdots \\
\end{smallmatrix} \right], [\mathbf{Z}] = \left[ \begin{smallmatrix}
    0 \\
    \vdots \\
    1 \\
    \vdots \\
\end{smallmatrix} \right], \text{ and }  [\mathbf{X}] = \left[ \begin{smallmatrix}
    X \\
    \vdots \\
    \alpha \cdot X +\beta \\
    \vdots \\
\end{smallmatrix} \right]  $$ 

Without loss of generality we can rewrite this as, %$\widetilde{[\mathbf{X}]} | [\mathbf{Z}] \sim \text{Multivariate-Normal} ([\mathbf{X}']  , \mathbf{1}) $
$$\widetilde{[\mathbf{X}]} | [\mathbf{Z}] \sim \text{Multivariate-Normal} ([\mathbf{X}']  , \mathbf{1}) $$
$$\text{where, }\widetilde{[\mathbf{X}]} = \left[ \begin{smallmatrix}
    \hat{X}_{1} \\
    \vdots \\
    {X}^*_{1} - \beta \\
    \vdots \\
\end{smallmatrix} \right], \text{ and }  [\mathbf{X}']  = \left[ \begin{smallmatrix}
    X \\
    \vdots \\
    \alpha \cdot X \\
    \vdots \\
\end{smallmatrix} \right] \qquad \qquad \qquad $$ 

From this and the fact that $X$ is random, we can recover the familiar form of Bayesian Linear regression as follows,
$$\widetilde{[\mathbf{X}]}  | [\mathbf{Z}] \sim \text{Multivariate-Normal} ([X_1, \hdots , X_{m+n}]^{\top}[\bm{\alpha}]  , \mathbf{1})$$
$$\text{here, } [\bm{\alpha}] = [1, \hdots , 1, \alpha, \hdots , \alpha]^{\top} \qquad \qquad\qquad$$

Now the Bayesian estimator of $X$ is the posterior mean (mode) of the above give as,

\begin{align*}
f_{\text{BE}}(\mathcal{D}) &= ( [\bm{\alpha}]^{\top}[\bm{\alpha}]+ \lambda_0  )^{-1} ([\bm{\alpha}]^{\top} \widetilde{[\mathbf{X}]} )\\
&= \frac{1}{(m+n\alpha^2 + \lambda_0) }\bigg(\sum_{i=1 }^{m} \hat{X}_i
 + \alpha \sum_{j=1 }^{n} {X_j^*} -n\alpha\beta \bigg) 
\end{align*}

\subsubsection{Properties}
\begin{align*}
\text{Now, } \mathbb{E}[f_{\text{BE}}] &= \dfrac{ (m + n\alpha^2)\mu }{(m+ n\alpha^2 + \lambda_0)} \approx \mu;  (\because \lambda_0<<m)\\
\text{and, } \text{Var}[f_{\text{BE}}] &=  \dfrac{ m \sigma^2 + n\alpha^2 \sigma_*^2 }{(m+ n\alpha^2 + \lambda_0)^2}
\end{align*}

\newpage
\subsection{Proof of Proposition \ref{prop:bevsce}}
\label{apx:bevsce}

\bevsce*

\noindent Proof. The Bias of the Bayesian Estimator ($f_{\text{BE}}$) is arbitrarily small and thus, in practice we consider the Bayesian Estimator as unbiased. Comparing their Mean Squared Errors of the Bayesian Estimator vs. the Conservative Estimator ($f_\text{CE}$), reduces to comparing their Variances.
%$$\text{Now, } \text{MSE}(f_{\text{BE}})&< \text{MSE}(f_{\text{CE}}) &\Longleftrightarrow \text{Var}[f_{\text{BE}}] <\text{Var}[f_{\text{CE}}]$$
\begin{alignat*}{3}
&\text{Now, } &\text{MSE}(f_{\text{BE}}) < \text{MSE}(f_{\text{CE}}) &\Longleftrightarrow \text{Var}[f_{\text{BE}}] <\text{Var}[f_{\text{CE}}]\\
&\Longleftrightarrow &   \dfrac{ m \sigma^2 + n\alpha^2 \sigma_*^2 }{(m+ n\alpha^2+ \lambda_0)^2} &< \frac{   \sigma^2}{ m} \\
&\Longleftrightarrow &    m^2 \sigma^2 + mn\alpha^2 \sigma_*^2  &<    \sigma^2 (m+ n\alpha^2 + \lambda_0)^2 \\
&\Longleftrightarrow &    m^2 \sigma^2 + mn\alpha^2 \sigma_*^2  &<    \sigma^2 (m+ n\alpha^2 )^2 \\
&\Longleftrightarrow &    m^2 \sigma^2 + mn\alpha^2 \sigma_*^2  &<    \sigma^2 (m^2+ n^2\alpha^4+ 2mn\alpha^2 )  \\
&\Longleftrightarrow &     mn\alpha^2 \sigma_*^2  &<    \sigma^2 ( n^2\alpha^4 + 2mn\alpha^2 )  \\
&\Longleftrightarrow &     m\sigma_*^2  &<    \sigma^2 ( n\alpha^2 + 2m\mu )  \\
&\Longleftrightarrow &     \frac{ \sigma_*^2}{ \sigma^2 }  &<    \frac{( n\alpha^2 + 2m\alpha^2 ) }{m} \\
&\Longleftrightarrow &     \frac{ \sigma_*^2}{ \sigma^2 }  &<   \bigg( {\frac{n}{m} + 2 } \bigg) \alpha^2.   
\end{alignat*}    
$\hfill \blacksquare$

\subsection{Proof of Proposition \ref{prop:noutility}}
\label{apx:bevsge}
\seven*

\noindent Proof Sketch. The Bias of the Bayesian Estimator ($f_{\text{BE}}$) is arbitrarily small and thus, in practice we consider the Bayesian Estimator as unbiased. Comparing their Mean Squared Errors of the Bayesian Estimator vs. the Greedy Estimator ($f_\text{GE}$), reduces to comparing their Variances.

\begin{alignat*}{3}
&\text{Now, } &\text{MSE}(f_{\text{BE}}) < \text{MSE}(f_{\text{GE}}) &\Longleftrightarrow \text{Var}[f_{\text{BE}}] <\text{Var}[f_{\text{GE}}]\\
&\Longleftrightarrow &   \dfrac{ m \sigma^2 + n\alpha^2 \sigma_*^2 }{(m+ n\alpha^2+ \lambda_0)^2} &<  \frac{  m \sigma^2  + \nicefrac{n\sigma_*^2}{\alpha^2}}{ (m+n)^{2}}  \\
&\Longleftrightarrow &   \dfrac{ m \sigma^2 + n\alpha^2 \sigma_*^2 }{(m+ n\alpha^2)^2} &<  \frac{  m \sigma^2  + \nicefrac{n\sigma_*^2}{\alpha^2}}{ (m+n)^{2}}  \\
&\Longleftrightarrow &   \dfrac{ m  + n\alpha^2 k }{(m+ n\alpha^2)^2} &<  \frac{  m   + \nicefrac{nk}{\alpha^2}}{ (m+n)^{2}}  \\
&\Longleftrightarrow &  (m+n)^2 (m+n\alpha^2k) &< (m+n\alpha^2)^2 (m + \frac{n}{\alpha^2} k)\\
&\Longleftrightarrow & n +2m + m\alpha^2k  +2\alpha^2kn  &< n\alpha^4 +2m\alpha^2 + \frac{mk}{\alpha^2} + 2nk\\
&\Longleftrightarrow & n\alpha^2 +2m\alpha^2 + m\alpha^4k  +2\alpha^4kn  &< n\alpha^6 +2m\alpha^4 + {mk} + 2nk\alpha^2\\
&\Longleftrightarrow & k (m\alpha^4 + 2\alpha^4n -m -2n\alpha^2) &< \alpha^2 (n\alpha^4 +2m\alpha^2 -n -2m)\\
&\Longleftrightarrow & k (m\alpha^4  -m + 2\alpha^4n -2n\alpha^2) &< \alpha^2 (n\alpha^4  -n+2m\alpha^2 -2m)\\
&\Longleftrightarrow & k (m (\alpha^4  -1) + 2n\alpha^2(\alpha^2 -1)) &< \alpha^2 (n(\alpha^4  -1)+2m (\alpha^2 -1))\\
&\Longleftrightarrow & k (m (\alpha^2  + 1) + 2n\alpha^2) &< \alpha^2 (n(\alpha^2  +1)+2m)\\
\end{alignat*}    
Now $k\leq2$. We set thus $k=2$ and solve the resulting stricter quadratic inequality in $\alpha^2$. The resulting set of feasible values for $\alpha$ would thus imply, $\text{MSE}(f_{\text{BE}}) < \text{MSE}(f_{\text{GE}})$. $ \hfill\blacksquare$

\section{Results in Section \ref{sec:lvbc} (Proposed Approach)}

\subsection{Parameter Inference and the derivation of the Evidence Lower Bound (ELBO)}
\label{apx:elbo}
To learn the parameters $\bm{\theta}$ of the model from observed analyst estimates and actual reported data $\mathcal{D}$, we want to maximize the likelihood, {\small $P(\mathcal{D}| \bm{\theta}) = P(\{\{\hat{X}_{ij}\}_{j=1}^{|\mathcal{A}|}\}_{i=1}^{\mathcal{|S|}}, \{X_i, \xi_i\}_{i=1}^{|\mathcal{S}|}|\{w_j\}_{j=1}^{|\mathcal{A}|}, \{ \alpha_k, \beta_k, \sigma_k  \}_{k=1}^{K})$}, which we can rewrite as follows: 
\begin{align*}  
P(\mathcal{D}| \bm{\theta})&= \prod^{|\mathcal{S}|}_{i=1} \prod_{j=1}^{|\mathcal{A}|} \sum_{k=1}^{K}  P_k(\hat{X}_{ij}|X_i, \xi_i, z_j) P(z_j|w_j)  \\
\log P(\mathcal{D}| \bm{\theta}) &= \sum^{|\mathcal{S}|}_{i=1} \sum_{j=1}^{|\mathcal{A}|} \log \sum_{k=1}^{K}  P_k(\hat{X}_{ij}|X_i, \xi_i, z_j) P(z_j|w_j)\\
 &= \sum^{|\mathcal{S}|}_{i=1} \sum_{j=1}^{|\mathcal{A}|} \log \mathbb{E}_{z_j|w_j} \big[  P_k(\hat{X}_{ij}|X_i, \xi_i, z_j) \big] \\
&\geq \sum^{|\mathcal{S}|}_{i=1} \sum_{j=1}^{|\mathcal{A}|}  \mathbb{E}_{z_j|w_j} \big[ \log P_k(\hat{X}_{ij}|X_i, \xi_i, z_j) \big] \quad \text{(From Jensen's Inequality)}\\
&\triangleq \textbf{ELBO}(\bm{\theta}; \mathcal{D})
\end{align*}

(Note the priors on the terms, $\{ \alpha_k, \beta_k, \sigma_k \}_{k=1}^{K}$, $P(\alpha,\beta, \gamma )$ is added to the $\textbf{ELBO}(\bm{\theta}; \mathcal{D})$ as defined above. ) 

\subsection{Choice of Hyperparameters and other experimental details}
\label{apx:exp}

For all the experiments we set the prior $\widetilde{\alpha}=1$, $\widetilde{\beta}=0$ and $\widetilde{\sigma}=2$. The prior regularization $\lambda$ was tuned between $\{1\times10^{2}, 1\times10^{3}, 1\times10^{4},  1\times10^{5} \}$ and minibatch size of 5000 for the \textbf{I/B/E/S} dataset; and $\lambda \in \{500, 750, 1000\}$ for the \textbf{CMU Delphi Flu Dataset} with no minibatching. 

\subsection{Posterior $X_i$ under the DAG model}
\label{apx:posterior}
\markovblanket*
\vspace{-1em}
 $$  X_i | \widetilde{[\mathbf{X}]}, \{z_j\}_{j=1}^{|\mathcal{A}|} \sim  \text{{Multivariate-Normal}}\Big( [\mathbf{X}]  \;  , [\bm{\Sigma}] \Big)$$
\begin{align*}
   \text{where,} \quad  [\mathbf{\Sigma}] &= \big(\lambda_0 + [\bm{\alpha}]^{\top}[\bm{\sigma}^2]^{-1}[\bm{\alpha}] \big)^{-1}, \\ [\mathbf{X}] &= [\mathbf{\Sigma}] ([\bm{\alpha}]^{\top} [\bm{\sigma}^2]^{-1} \widetilde{[\mathbf{X}]}) \quad  \text{and, }
\end{align*}
{  $$ {{ \widetilde{[\mathbf{X}]} =  \left[ \begin{smallmatrix}
    \hat{X}_{i0} - \beta_{z_0}       \\
    \hat{X}_{i1} - \beta_{z_1}      \\
    \vdots \\
    \hat{X}_{i|\mathcal{A}|} - \beta_{z_{|\mathcal{A}|} }             \\
\end{smallmatrix} \right]}},
[\bm{\alpha}] =\left[ \begin{smallmatrix}
 \alpha_{z_0}\\ \alpha_{z_1}\\ \vdots \\ \alpha_{z_{|\mathcal{A}|}}  \end{smallmatrix} \right],
 [\bm{\sigma}^2] = \text{diag}
 \left[ \begin{smallmatrix}
 \sigma^2_{z_0}\\ \sigma^2_{z_1}\\ \vdots \\ \sigma^2_{z_{|\mathcal{A}|}}  \end{smallmatrix} \right]$$}
\vspace{-1em}
{\small Proof Sketch. Place a weak conjugate prior on
{ $X_i \sim \mathcal{N} (0,  \nicefrac{1}{\lambda_0^2})$}}.\\
{\small \begin{align*}
\hat{X}_{ij} | X_i, z_j &\sim \mathcal{N} (\alpha_{z_j} \! \! \cdot \!  X_i + \beta_{z_j} \;  , \;   \sigma_{z_j}^{2})\\
\hat{X}_{ij} -\beta_{z_j} | X_i, z_j &\sim \mathcal{N} (\alpha_{z_j} \! \! \cdot \!  X_i  \;  , \;   \sigma_{z_j}^{2})\\
\{ \hat{X}_{ij} - \beta_{z_j} \}_{j=1}^{|\mathcal{A}|}| X_i,   \{z_j\}_{j=1}^{|\mathcal{A}|} &\sim \prod_{j=1}^{|\mathcal{A}|} \mathcal{N}(\alpha_{z_j} \! \! \cdot X_i \;  , \; \sigma_{z_j}^{2})
\end{align*}
Rewriting in matrix form as in , we get
{$$\widetilde{[\mathbf{X}]} | X_i,   \{z_j\}_{j=1}^{|\mathcal{A}|} \sim  \text{{Multivariate-Normal}}\Big(X_i^{\top} [\bm{\alpha}]  \;  , [\bm{\sigma}^2] \Big)\label{eqn:mvn}$$}}

\noindent We thus arrive at a conditional distribution similar to the one derived for the Bayesian Estimator in Section \ref{sec:estimators}. From Equation \ref{eqn:mvn} and \cite{cepeda2000bayesian, cepeda2005bayesian} pertaining to Bayesian Linear Regression under Heteroscedasticity, we arrive at the posterior. $ \hfill \blacksquare$

% The generating process under these assumptions can then be written as, 

% $$\hat{X}_{i} | X, z_i &\sim \mathcal{N} (\alpha_{z_i} \! \! \cdot \!  X_i + \beta_{z_i} \;  , \;   \sigma_{z_i}^{2})$$
% $$\hat{X}_{ij} -\beta_{z_j} | X_i, z_j &\sim \mathcal{N} (\alpha_{z_j} \! \! \cdot \!  X_i  \;  , \;   \sigma_{z_j}^{2})$$
% $$\{ \hat{X}_{ij} - \beta_{z_j} \}_{j=1}^{|\mathcal{A}|}| X_i,   \{z_j\}_{j=1}^{|\mathcal{A}|} &\sim \prod_{j=1}^{|\mathcal{A}|} \mathcal{N}(\alpha_{z_j} \! \! \cdot X_i \;  , \; \sigma_{z_j}^{2})$$

\newpage
\subsection{Gibbs Sampler for Inverse Inference}\label{apx:gibbssampler}
\begin{algorithm}[!htbp]
{
\KwIn{$\{\{\hat{X}_{ij}\}_{i=1}^{|\mathcal{S}|}\}_{j=1}^{|\mathcal{A}|}, \{\alpha_k, \beta_k, \sigma_k \}_{k=1}^{K},\{ w_j \}_{j=1}^{\mathcal{|A|}} $}
\textbf{Initialize:}\\
\For{$i\leftarrow 1$ \KwTo $|\mathcal{S}|$}{
%\For{$j\leftarrow 1$ \KwTo $\mathcal{A}$}{
$X_i^{(0)}$ =  $\frac{1}{|\mathcal{A}|} \sum_j  \hat{X}_{ij}$ ; { (Consensus Estimate)}\\
 $\xi_i^{(0)}$ =  $\mathbbm{1}\{ X_{i}^{(0)} > 0  \} $;\\

}

\For{$n\leftarrow 1$ \KwTo $N$}{
\For{$j\leftarrow 1$ \KwTo $\mathcal{A}$}{
$z_j^{(n)} \sim \text{Discrete}\bigg(\textsc{Soft-Max}(w_j)\bigg)$;\\
}
\For{$i\leftarrow 1$ \KwTo $\mathcal{S}$}{$X_i^{(n)} \sim P_{\text{posterior}} (X_i^{(n)} | \{ z_j^{(n)} \}_{j=1}^{\mathcal{|A|}}, \xi_i^{(n-1)},  \{ \alpha_k, \beta_k, \sigma_k \}_{k=1}^{K} )$; (As in Prop. \ref{prop:posterior})\\
$\xi_i^{(n)} = \mathbbm{1}\{ X_i^{(n)} >0 \} $;
}}
\KwOut{$ \{ X_i^{(0)},  X_i^{(1)},\cdots, X_i^{(1)}  \}$}    
\caption{ Gibbs Sampler for $P(\{X_i, \xi_i \}_{i=1}^{\mathcal{|S|}}| \cdot )$}\label{alg.mainLoop}}
\end{algorithm}

\newpage

\section{Results in Tabular format}
\label{apx:results}
Results from experiments in numerical form are in the tables above.
\begin{table*}[!htbp]
 \centerline{
 \begin{tabular}{r|c|c|c}
        \toprule \midrule
        &\multicolumn{3}{c}{\textsc{Micro Average}}\\ \midrule
         \textsc{Model}& \textbf{RMSE} &  \textbf{MAE} & \textit{\textbf{R}}$^\mathbf{2}$   \\ \midrule
        \textbf{NE}&$0.3303\pm0.001$&$0.2085\pm0.001$&$0.1482\pm0.007$\\
        \textbf{WE}&$0.3384\pm0.002$&$0.1932\pm0.001$&$0.128\pm0.006$\\
        \textbf{RE}&$0.3618\pm0.002$&$0.2181\pm0.001$&$0.0264\pm0.001$\\
        \textbf{BRE}&$0.3278\pm0.001$&$0.2046\pm0.001$&$0.1627\pm0.007$\\
 \midrule
\rowcolor{Gray}         \textbf{LVBC}&$0.3255\pm0.001$&$0.2039\pm0.001$&$0.1738\pm0.007$\\
 \midrule \bottomrule 
    \end{tabular}}
    \caption{Results on the Weekly National \textbf{(ILI)} Forecasting Estimates} 
    \label{tab:FY2}
\end{table*}

\begin{table*}[!htbp]
    \centerline{
    \begin{tabular}{r|c|c|c|c|c}
        \toprule \midrule
        &\multicolumn{2}{c|}{\textsc{Macro Average}}&\multicolumn{3}{c}{\textsc{Micro Average}}\\ \midrule
         \textsc{Model}& \textbf{RMSE} &  \textbf{MAE} &\textbf{RMSE} &  \textbf{MAE} & \textit{\textbf{R}}$^\mathbf{2}$   \\ \midrule
         \textbf{NE}&$0.4890\pm0.001$&$0.4119\pm0.001$&$0.7090\pm0.001$&$0.4113\pm0.001$&$0.7776\pm0.001$\\
\textbf{WE}&$0.5069\pm0.001$&$0.4274\pm0.001$&$0.7288\pm0.001$&$0.4264\pm0.001$&$0.7650\pm0.002$\\
\textbf{RE}&$0.5058\pm0.001$&$0.4293\pm0.001$&$0.7266\pm0.001$&$0.4292\pm0.001$&$0.7670\pm0.001$\\
\textbf{BRE}&$0.4866\pm0.001$&$0.4095\pm0.001$&$0.7077\pm0.001$&$0.4085\pm0.001$&$0.7782\pm0.001$\\ \midrule
          \rowcolor{Gray} \textbf{LVBC}&$0.4866\pm0.001$&$0.4098\pm0.001$&$0.7065\pm0.001$&$0.4085\pm0.001$&$0.7789\pm0.001$\\ \midrule \bottomrule 
    \end{tabular}}
    \caption{Results on the Fiscal Year 1 (\textbf{FY1}) Year Estimates} 
    \label{tab:FY1}
\end{table*}

\begin{table*}[!htbp]
 \centerline{
 \begin{tabular}{r|c|c|c|c|c}
        \toprule \midrule
        &\multicolumn{2}{c|}{\textsc{Macro Average}}&\multicolumn{3}{c}{\textsc{Micro Average}}\\ \midrule
         \textsc{Model}& \textbf{\textsc{RMSE}} &  \textbf{MAE} &\textbf{RMSE} &  \textbf{MAE} & \textit{\textbf{R}}$^\mathbf{2}$   \\ \midrule
        \textbf{NE}&$1.0159\pm0.002$&$0.8631\pm0.002$&$1.6227\pm0.005$&$0.8571\pm0.002$&$0.4676\pm0.002$\\
        \textbf{WE}&$1.0127\pm0.002$&$0.8587\pm0.002$&$1.6187\pm0.005$&$0.8530\pm0.002$&$0.4703\pm0.002$\\
        \textbf{RE}&$1.0127\pm0.002$&$0.8587\pm0.002$&$1.7268\pm0.005$&$0.9328\pm0.002$&$0.3992\pm0.002$\\
        \textbf{BRE}&$1.0082\pm0.002$&$0.8506\pm0.002$&$1.5877\pm0.005$&$0.8451\pm0.001$&$0.4888\pm0.002$\\ \midrule
\rowcolor{Gray} \textbf{LVBC}&$1.0096\pm0.002$&$0.8514\pm0.002$&$1.5825\pm0.005$&$0.8459\pm0.001$&$0.4919\pm0.002$\\ \midrule \bottomrule 
    \end{tabular}}
    \caption{Results on the Fiscal Year 2 (\textbf{FY2}) Year Estimates} 
    \label{tab:FY2}
\end{table*}

\end{document}